\documentclass[fleqn,usenatbib]{mnras}

\usepackage{newtxtext,newtxmath,pdflscape}

\usepackage[T1]{fontenc}
\usepackage{xcolor}

\DeclareRobustCommand{\VAN}[3]{#2}
\let\VANthebibliography\thebibliography
\def\thebibliography{\DeclareRobustCommand{\VAN}[3]{##3}\VANthebibliography}

\usepackage{graphicx}	
\usepackage{amsmath}	

\usepackage{tabularx}

\title[Photometric SMBH mass estimates in AGN]{Exploring mass measurements of supermassive black holes in AGN using GAMA photometry and spectroscopy}

\author[S. Casura, D. Ili\'c,  et al.]{
Sarah Casura$^{1}$\thanks{The first two authors should be regarded as Joint First Authors.}\thanks{E-mails: dragana.ilic@matf.bg.ac.rs (DI), sarah.casura@uni-hamburg.de (SC)},
Dragana Ili\'c$^{1,2\star}$,
Jonathan Targaczewski$^{1}$,
Nemanja Raki\'c$^{3}$
and Jochen Liske$^{1}$
\\
$^{1}$Hamburger Sternwarte, Universit{\"a}t
Hamburg, Gojenbergsweg 112, 21029 Hamburg, Germany\\
$^{2}$Department of Astronomy, University of Belgrade - Faculty of Mathematics, Studentski trg 16, 11000 Belgrade, Serbia\\
$^{3}$Physics Department, Faculty of Natural Sciences and Mathematics, University of Banjaluka, Mladena Stojanovi\'ca 2, 78000 Banjaluka, RS, \\ Bosnia and Herzegovina
}

\date{Accepted XXX. Received YYY; in original form ZZZ}

\pubyear{2024}

\begin{document}
\label{firstpage}
\pagerange{\pageref{firstpage}--\pageref{lastpage}}
\maketitle

\begin{abstract}
In the era of massive photometric surveys, we explore several approaches to estimate the masses of supermassive black holes (SMBHs) in active galactic nuclei (AGN) from optical ground-based imaging, in each case comparing to the independent SMBH mass measurement obtained from spectroscopic data. We select a case-study sample of 28 type 1 AGN hosted by nearby galaxies from the Galaxy And Mass Assembly (GAMA) survey. We perform multi-component spectral decomposition, extract the AGN component and calculate the SMBH mass from the broad H$\alpha$ emission line width and luminosity. The photometric $g$ and $i$ band data is decomposed into AGN+spheroid(+disc)(+bar) components with careful surface brightness fitting. From these, the SMBH mass is estimated using its relation with the spheroid S\'ersic index or effective radius (both used for the first time on ground-based optical imaging of AGN); and the more widely used scaling relations based on bulge or galaxy stellar mass. 
We find no correlation between the H$\alpha$-derived SMBH masses and those based on the spheroid S\'ersic index or effective radius, despite these being the most direct methods involving only one scaling relation. The bulge or galaxy stellar mass based methods both yield significant correlations, although with considerable scatter and, in the latter case, a systematic offset. We provide possible explanations for this and discuss the requirements, advantages and drawbacks of each method. These considerations will be useful to optimise stategies for upcoming high quality ground-based and space-borne sky surveys to estimate SMBH masses in large numbers of AGN.
\end{abstract}





\section{Introduction}
\label{sec:intro}

Most galaxies, if not all, host a super-massive black hole (SMBH) in their centre \citep[see][for a review]{2013ARA&A..51..511K}. A current important task and an ongoing challenge is to accurately measure the mass of these SMBH, using either direct or indirect methods \citep{peterson14}, where the latter are methods that give the SMBH mass from so-called scaling relations. Many different empirical scaling relations, i.e. correlations between different observables of the host galaxy and the SMBH within have been established \citep[see e.g.][]{2000ApJ...539L..13G,2001AIPC..586..363K,2015ApJ...802...14S}, suggesting a tight connection and possible co-evolution of the SMBH and the host galaxy \citep{2013ARA&A..51..511K}. These scaling relations are essential for our understanding of the co-evolution of SMBH and galaxies across cosmic times.

Such established scaling relations include for example the well known relation between the mass of the SMBH and the velocity dispersion of the stars within the bulge \citep{1995ARA&A..33..581K, 2000ApJ...539L...9F, 2023arXiv230801800C} or the relation between the SMBH mass and the bulge mass \citep{2003ApJ...589L..21M}. With these relations one can estimate the mass of the SMBH by using the more readily available data on the host galaxy. This is especially useful for distant galaxies where the inner region of the galaxies, which is under the gravitational influence of the SMBH, cannot be resolved. 

Particularly interesting is to measure the SMBH mass in active galactic nuclei (AGN), as these object are among the brightest objects in the universe and can be detected at the largest cosmological distances, even within the first billion years after the big bang \citep{2021ApJ...907L...1W}. For AGN, there are both direct and indirect methods for SMBH mass estimates based mainly on the broad emission lines and the assumption that the gas from which these lines originate (i.e., the broad line region - BLR) is virialized \citep[for a review, see][]{pop20}, but novel methods based on narrow emission lines are also proposed \citep{2019MNRAS.487.3404B}. Additionally, there were some suggestions to use the Fundamental Plane of black hole accretion, an empirical correlation between the SMBH mass, radio luminosity, and X-ray luminosity, to obtain the SMBH masses in AGN \citep[see][and references therein]{2022MNRAS.516.6123G}, which could be used for a wide range of SMBH masses and accretion rates. For getting a reliable estimate of the SMBH mass from indirect methods, one should preferably use several independent methods that are based on different observables and scaling relations. On the other hand, independent methods could be used for calibration of direct or indirect methods \citep{peterson14}, as for example to constrain the virial factor $f$ used in the virial theorem using the observed BLR gas velocities and its characteristic dimensions \citep{1977SvAL....3....1D, 1986ApJ...305..175G}.

Most of the methods for SMBH mass are based on spectroscopy, i.e., either on stellar velocity dispersion or BLR gas velocities measured from spectral lines \citep{2005SSRv..116..523F,Greene05}.  For large samples of AGN, typically the SMBH mass is estimated using the radius-luminosity scaling relation that connects the luminosity of the continuum (or luminosity of the broad H$\alpha$ or H$\beta$ lines) and the size of the BLR \citep[see e.g.][etc.]{kaspi2000, Bonta2020, Xiao_2011}. 
Alternatively, and also in the absence of spectroscopic data, the SMBH mass can be inferred from its tight correlation with the stellar mass in the bulge, or - with larger scatter - with galaxy stellar mass \citep[e.g.][]{2013ARA&A..51..511K, Bentz2018}. Similar relations between the SMBH mass and bulge or galaxy luminosity have also been established, where the near-infrared K-band is preferred over optical wavelengths in order to minimise the influence of young stellar populations and dust attenuation \citep[and references therein]{2013ARA&A..51..511K}. Further properties of the host galaxy, such as the \citet{Sersic1963} index $n$ of its spheroid (bulge) have also been found to correlate with SMBH mass \citep{Graham_2001, Graham_2003, Graham_2007, 2016ApJ...821...88S}, with the recent update in \citet{Sahu_2020} devoted to deriving different relations for early-type galaxies (ETGs) and late-type galaxies (LTGs). 

These relations have the advantage that the SMBH mass can be estimated from photometric data alone, which is relatively cheap to acquire in terms of telescope time compared to spectroscopy. However, a robust photometric estimate of stellar mass in the bulge typically requires multi-wavelength data to derive an appropriate mass-to-light ratio \citep[M/L; usually involving another scaling relation - e.g.][]{Bell2001, Into2013} and a careful analysis to separate the bulge or spheroid from other galaxy components \citep{Bentz2018, Sahu_2020}. This becomes even more challenging in the presence of strong AGN emission \citep{bentz09, Bentz2018, Zhuang2023}. Using the total galaxy stellar mass instead of the bulge mass circumvents the need to isolate the spheroid component, but comes at the cost of larger intrinsic scatter and possible AGN contamination if not accounted for; while using relations involving galaxy or bulge luminosity instead of mass attempts to reduce the number of separate scaling relations applied, which, however, results in less universally applicable results since they are strongly wavelength-dependent \citep{2013ARA&A..51..511K, Bentz2018}.

Using the M$_{\mathrm{BH}}$ - $n$ relation has the advantages that it requires only single-band images (which may also be photometrically uncalibrated) and that estimates of $n$ do not depend on galaxy distance or an uncertain M/L ratio, while still being largely wavelength-independent and ``direct'' in the sense that only a single scaling law is needed \citep{Graham_2003}. However, this relation has never before been used in a study dedicated to AGN only, primarily due to many challenges in surface brightness disentangling of strong AGN emission from the spheroid component \citep[e.g.][]{bentz09, Bentz2018, Zhao2021, Zhuang2023}, especially when using ground-based imaging alone. In fact, even in the absence of an additional AGN component, and for galaxies in the local universe only, disentangling the main stellar components (bulges and discs) is often challenging \citep[see relevant discussions in][to name just a few]{Allen2006, Gadotti2009, Simard2011, Lackner2012, Kennedy2016, Kim2016, Meert2016, Bottrell2019, Casura2022, Haeussler2022, Robotham2022}. Thus, obtaining reliable S\'ersic indices for potentially barely resolved bulges in ground-based imaging is difficult on its own, and is further complicated by the presence of an often highly dominant AGN.

Motivated by upcoming vast and deep ground-based and spaceborne surveys such as the Vera C. Rubin Observatory Legacy Survey of Space and Time \citep[LSST;][]{Ivezic2019}\footnote{\url{https://rubinobservatory.org}} or the Euclid mission \citep{2022A&A...662A.112E}\footnote{\url{https://sci.esa.int/web/euclid/-/summary}}, in this paper we focus on exploring the possibility to predict the SMBH mass in AGN using photometry only. Therefore in this work we study a case-study sample of AGN from the Galaxy And Mass Assembly \citep[GAMA;][]{2009A&G....50e..12D}, for which we have both spectral and photometric data. We focus on the type 1 AGN, which are those with broad emission lines in their spectra \citep[e.g.,][]{netzer15}, located in nearby galaxies for which extended structures are visible. We opt to use GAMA data since it has both spectroscopic and state-of-the-art ground-based imaging data available, the latter of which have already been analysed in the context of the bulge-disc decomposition study of \citet{Casura2022}. We build upon this work to perform detailed multi-component surface brightness decompositions, in order to separate the AGN from other galaxy components. This allows us to extract spheroid S\'ersic indices and effective radii, which we convert to black hole masses via the \citet{Sahu_2020} relations. For comparison, we also compute black hole masses with the galaxy mass - SMBH mass scaling relation and the bulge mass - SMBH mass scaling relation, where we use the relations for active galaxies presented in \citet{Bentz2018}. The stellar mass is estimated from the $g - i$ colour of the entire galaxy or the spheroid component, respectively, using the M/L derived in \citet{Taylor2011}. In all cases, we compare to our ``reference'' SMBH masses obtained from a spectral analysis and the careful extraction of the pure broad emission line parameters, which is described in \cite{2023ApJS..267...19I} in order to get the SMBH mass using the scaling relations based on the broad emission lines. For all methods, we discuss advantages and weaknesses, as well as possibilities and recommendations for future photometric surveys.

This paper is structured as follows: in Section~\ref{sec:datasample} we briefly describe the GAMA data and our sample selection, whereas Section~\ref{sec:scalingrelations} presents the various scaling relations used in deriving SMBH mass estimates. In Section~\ref{sec:dataanalysis} we introduce the  multi-component spectral and surface brightness decomposition and describe how we derived SMBH masses in each case. In Section~\ref{sec:results} we present and discuss the results of the spectral and photometric fitting, comparing the different SMBH mass estimates. Finally, Section~\ref{sec:conclusion} outlines the conclusions and future prospects. 

\section{Data and sample selection}
\label{sec:datasample}

Our sample comes from the GAMA survey\footnote{\url{https://www.gama-survey.org/}} 
which contains $\sim$\,$300,000$ galaxies over an area of $\sim 286 \,\text{deg}^2$ split into three equatorial and two southern sky survey regions \citep{Liske_2015}. 

The GAMA survey spectroscopy is highly complete to a magnitude of $r<19.8\,\text{mag}$, using a combination of data from the Sloan Digital Sky Survey \citep[SDSS,][]{2008ApJS..175..297A}, Millennium Galaxy Catalogue \citep[MGC,][]{2003MNRAS.344..307L} the 2dF Galaxy redshift survey \citep[2dFGRS,][]{2001MNRAS.328.1039C} and a dedicated campaign using the AAOmega spectrograph on the Anglo-Australian Telescope \citep[AAT;][]{Liske_2015}. The spectra provided by the GAMA team were taken in the blue and red arms separately, and later spliced together at $5700\,$\AA \, \citep{Baldry_2014}, to yield a continuous wavelength coverage of 3720–8850\,\AA\ at a resolution of $\approx$\,3.5\,\AA\, in the blue and $\approx$\,5.5\,\AA \, in the red channel. The GAMA team measured the redshift using the \textsc{autoz} \citep{Baldry_2014} and \textsc{runz} codes. In general, the \textsc{autoz} redshifts are preferred, since \textsc{runz} requires human verification of the found redshift. However it contains quasar template spectra in contrast to \textsc{autoz}, thus here we use \textsc{runz} redshifts.\\

On the imaging side, the GAMA database includes data from a number of independent imaging survey teams with coordinated survey regions and negotiated data sharing agreements. In this analysis, we use $i$-band data from DR4.0 of the Kilo-Degree Survey \citep[KiDS,][]{Kuijken2019}. KiDS is a wide-field imaging survey in the Southern sky using the VLT Survey Telescope (VST) at the ESO Paranal observatory. A total of 1350\,deg$^2$ are mapped in the optical broad-band filters $u$, $g$, $r$, $i$; including the GAMA II equatorial survey regions (fully covered as of DR3.0). The `science' tiles provided on the public KiDS database\footnote{\url{http://kids.strw.leidenuniv.nl/DR4/index.php}} are composed of 5 co-added dithers taken in immediate succession and re-gridded onto a common pixel scale of 0.2\,arcsec. Associated weight (inverse variance) maps and masks are also provided. For details, see \citet{Kuijken2019}. 

Here, we choose the $i$-band data as a reasonable trade-off between the data quality in terms of depth and seeing (which is best in the KiDS $r$-band; \citealt{Kuijken2019}) and the prominence of the bulge component, the transparency of interstellar dust and the contrast between the AGN and the host, all of which increase as a function of wavelength \citep[e.g.,][]{Popescu2011, Zhao2021}. The $i$-band has a limiting magnitude of 23.7\,mag (5$\sigma$ in a 2\,arcsec aperture) and a point spread function (PSF) size of typically $\sim$0.9\,arcsec. At the redshifts of our sample ($z < 0.1$), this corresponds to a physical resolution of 1-2\,kpc. This is comparable to the sizes of the spheroids found (with effective radii ranging from 0.6 to 6\,kpc), thus we are working at the resolution limit of the data. For two of the methods involving stellar mass estimates (cf. Section~\ref{sec:errorestimation}) we additionally use the $g$-band data to obtain $g - i$ colours. The $g$-band limiting magnitude is 25.1\,mag and the seeing approximately 0.9\,arcsec, consistent with that in the $i$-band.\\

In addition to the basic spectroscopic and imaging data, the GAMA database contains many derived data products on specific scientific topics, structured into data management units (DMUs). We use the bulge-disc decompositions in the \textsc{BDDecomp} DMU \citep{Casura2022} as a starting point for the multi-component surface brightness decompositions performed in this work, see Section~\ref{sec:sbfitting} for details.
To determine the number of components of a galaxy, we make use of the visual morphological classifications from the \textsc{gkvMorphologyv02} and \textsc{VisualMorphologyv03} DMUs, which we describe in more detail in Section~\ref{sec:visualgalaxyclassification}. To convert apparent magnitudes and spectral fluxes into absolute magnitudes and luminosities, we use the distance moduli from the \textsc{DistancesFramesv14} catalogue in the \textsc{LocalFlowCorrection} DMU \citep[originally described by][]{Baldry2012}, choosing the values for a `737' cosmology (H$_0=70$ km s$^{-1}$ Mpc$^{-1}$, $\Omega_m = 0.3$ and $\Omega_{\Lambda} = 0.7$). To correct the broad-band magnitudes and colours for Galactic extinction, we apply the Galactic extinction values from the \textsc{GalacticExtinctionv03} catalogue in the \textsc{EqInputCat} DMU \citep{Baldry2010}.\\

Although the GAMA survey was primarily motivated to study the evolution and formation of galaxies, there is a fraction of nearby and extended AGN within the main sample \citep{Gordon_2017, Gordon_2018}.

Type 1 AGN were identified following the selection strategy from \citet{Gordon_2017} that makes use of the GAMA DMUs \textsc{SpecLineSFRv05}\footnote{\url{https://www.gama-survey.org/dr4/schema/dmu.php?id=104}} and \textsc{SpecCatv27}\footnote{\url{https://www.gama-survey.org/dr4/schema/dmu.php?id=91}} that contain complete information on the spectra together with the spectroscopic redshifts derived from \textsc{autoz} and \textsc{runz} codes. The DMU SpecLineSFR provides measurements on emission and absorption lines, within which the table GaussFitComplexv05 contains the result from the fits performed with two Gaussian components for the H$\alpha$ and H$\beta$ lines to account for the narrow and broad component respectively \citep[see for details][]{Gordon_2017}. 
	
The sample was selected by first of all only considering spectra from the SDSS or taken by the GAMA team. Further, we require an average spectral signal to noise ratio (S/N) of $3$ or higher, as well as a redshift $z<0.1$ with a normalized redshift quality (NQ) of $\geq 3$, indicating that the redshift is suitable for scientific purposes \citep{Liske_2015}. The spectral lines of H$\alpha$ and H$\beta$ are required to have a broad component with the measured Full Width at Half Maximum (FWHM) of $1000\,\text{km/s}$ or greater, found within \textsc{GaussFitComplexv05} table. Thus, the selection criteria used are: 
(i) $z<0.1$, (ii) $\text{NQ}\geq 3$, (iii) $\text{S/N}\geq3$, (iv) FWHM(${\text{H}\alpha}) > 1000\,\text{km s}^{-1}$, (v) FWHM(${\text{H}\beta}) > 1000\,\text{km s}^{-1}$.

This selection resulted in a sample of $155$ objects in total. Since the automated spectral fitting is optimised for galaxies without AGN, we then visually inspected all objects and discarded spectra which showed signs of fringing artifacts or had no broad line component for the H$\alpha$ and H$\beta$ line, leaving a sample of $35$ objects with clear signatures of broad emission lines. An example of a spectrum from the sample can be seen in Figure \ref{fig: spectra} (top panel, grey solid line). The S/N ratio of this sample is in the range $10$ -- $80$, with a mean of~$\sim$$20$.

Since we also need imaging data for the photometric decompositions, we further limited our sample to the GAMA II equatorial survey regions which are covered by KiDS, removing another 7 objects from the sample. This leaves a total of 28 objects for analysis. Most of the spectra for our sample originate from the SDSS ($26$), while for $2$ objects the spectral data were taken by the GAMA team. 

It is important to acknowledge that our sample of $28$ AGN is far from being complete or representative, and thus it is unsuitable for any statistical analysis of the type 1 AGN population. However, it contains a variety of type 1 AGN of different spectral properties and thus physical properties, i.e., broad emission lines of different widths and strengths which implies different SMBH masses and accretion rates. The host galaxies of the selected sample show diverse properties as well, from elliptical and lenticular galaxies to spirals with and without bars, sometimes with nearby stars or interacting galaxies. Considering all of the above, the selected sample is appropriate for the case study presented in this work.

We emphasize that the emission line parameters provided in the table \textsc{GaussfitComplexv05} are extracted by fitting individually the line of interest, assuming a linear underlying continuum. This introduces a certain amount of uncertainty when estimating the AGN continuum level, leading sometimes to false detection of the broad component, especially for noisy data. Note that the procedures used by \cite{Gordon_2017} do not account for the contribution of the host galaxy, or of the Fe II multiplets, which are both considered in this work.

In addition, in type 1 AGN the broad spectral lines are typically complex and cannot be described with a single broad Gaussian \citep[see e.g.,][]{pop04}. Therefore, in this work we performed multicomponent spectral decomposition of the full optical spectrum, which fits the H$\alpha$ and H$\beta$ lines simultaneously with the underlying continuum and narrow (such as [O III] and [N II] doublets) and satellite lines (such as broad He II or complex Fe II emission). We are aiming to extract the contribution of the AGN emitting component with greater care than in the automated survey manner (described in Section~\ref{sec:3.1}).

The full sample is given in Table \ref{tab:sbfittingresults}, in which the first 4 columns list the following: the unique identifier for the object in this work, the GAMA CATAID, the redshift $z$ for the object found by the runz code, and the galaxy type, which is derived in Section~\ref{sec:visualgalaxyclassification}. Further columns in Table \ref{tab:sbfittingresults} will be discussed within Section \ref{sec:results}.

\section{Scaling relations for SMBH mass estimates}
\label{sec:scalingrelations}
The mass of the SMBH can be estimated from the physical properties of the BLR, namely its size and gas velocity, assuming that the BLR is gravitationaly bound to the SMBH, so that the virial theorem can be applied. This gives that the mass M$_{\rm BH}$ is:
	\begin{equation*}
	\mathrm{M}_{\rm BH} = f\frac{R_{\rm BLR}v^2}{G}\,\text{,}
	\label{eqn:virial}
	\end{equation*} 
where $R_{\rm BLR}$ is the radius at which the BLR resides from the SMBH, $v$ is the velocity of the BLR, $G$ is the gravitational constant, and $f$ is a dimensionless parameter which accounts for the inclination and kinematics of the AGN. This method was first used by \cite{1977SvAL....3....1D}, which is why it is sometimes referred to as the Dibai method \citep{2006ApJ...648..128K, 2011sf2a.conf..577G}. 
Over the years this method has been widely used to estimate the SMBH mass, if the size of the BLR and the velocity of the gas within are measured. 
The velocity can easily be measured from the width of the spectral line, whereas the radius $R_\text{BLR}$ may be calculated from a radius-luminosity scaling relation that connects the luminosity of the continuum and the size of the BLR \citep[see e.g.,][]{kaspi2000, bentz09}, leading to the scaling relations for the SMBH mass estimates \citep[for a recent review see][]{pop20}. 

Usually the AGN continuum luminosity at 5100 \AA\, is used to predict the BLR size, however it has been shown that one may use the luminosity of the broad H$\alpha$ or H$\beta$ lines instead without any loss of precision \citep[see e.g.,][]{2007ApJ...670...92G, 2012ApJ...755..167D, 2014JKAS...47..167W, Bonta2020}, which minimize the effects of the host-galaxy contribution or from non-thermal jet emission to the observed continuum luminosity \citep{Greene05}. We decide to use the H$\alpha$ line rather than H$\beta$ mainly since the majority of the objects in our sample are low-luminosity AGN in which the H$\beta$ line is under much higher influence of the host galaxy and noise effects. We note also that spectra within the GAMA survey taken with the AAOmega spectrograph are divided into two ranges spliced together at $5700\,$\AA, where the blue part is of lower quality in terms of S/N ratio \citep{Baldry_2014}. In addition, the H$\beta$ line is under larger influence of satellite and narrow emission line, posing additional sources of uncertainty when extracting from noisy spectra. Thus, we decided to extract spectral parameters (flux and line width) of the broad H$\alpha$ line, since the uncertainty on the
H$\beta$ line or nearby continuum luminosity would be higher.

In this work, for the SMBH mass M$_{\text{BH}}$ (in units of M$_\odot$) we use the relation of \cite[their equation 6]{Xiao_2011} that makes use of the luminosity and FWHM of the H$\alpha$ line:
	\begin{eqnarray}
		\text{log}(\mathrm{M}_{\text{BH}}) &=& 6.40^{+0.09}_{-0.07} + (0.45\pm0.05) \text{log}\biggl(\frac{L({\text{H}\alpha})}{10^{42}\,\text{erg s}^{-1}}\biggr) \nonumber\\
				&+& (2.06\pm0.06)\text{log}\biggl(\frac{\text{FWHM(H}\alpha\text{)}}{10^3\,\text{km s}^{-1}}\biggr)\,\text{,}
		\label{eqn:xiao}
	\end{eqnarray}
where $L({\text{H}\alpha})$ is the luminosity of the H$\alpha$ line in erg/s, FWHM$(\text{H}\alpha)$ is the FWHM of the H$\alpha$ line in km/s. \\

The spheroid S\'ersic index, $n$, obtained from the surface brightness fitting is converted to an estimate of the SMBH mass according to the relations presented in \citet[their equations~5 and~6]{Sahu_2020}: 
\begin{equation}
    \log_{10}(\mathrm{M}_{\mathrm{BH}} / \mathrm{M}_{\odot}) = (3.95 \pm 0.34) \log_{10}(n / 3) + (8.15 \pm 0.08) 
	\label{eq:SahurelationETG}
\end{equation}
for ETGs with a root-mean-squared (rms) scatter of 0.65\,dex; and 
\begin{equation}
    \log_{10}(\mathrm{M}_{\mathrm{BH}} / \mathrm{M}_{\odot}) = (2.85 \pm 0.31) \log_{10}(n / 3) + (7.90 \pm 0.14) 
	\label{eq:SahurelationLTG}
\end{equation}
for LTGs (rms scatter 0.67\,dex). 

Similarly, the spheroid effective radius $R_\mathrm{e}$ is used to obtain a SMBH mass from the relations presented in table~1 of \citet{Sahu_2020}: 
\begin{equation}
    \log_{10}(\mathrm{M}_{\mathrm{BH}} / \mathrm{M}_{\odot}) = (2.13 \pm 0.22) \log_{10}(R_\mathrm{e}) + (8.34 \pm 0.09) 
	\label{eq:SahuReETGwithdisc}
\end{equation}
for ETGs with a disc (i.e. S0 galaxies; rms scatter 0.55\,dex);
\begin{equation}
    \log_{10}(\mathrm{M}_{\mathrm{BH}} / \mathrm{M}_{\odot}) = (1.78 \pm 0.24) \log_{10}(R_\mathrm{e}) + (7.24 \pm 0.25) 
	\label{eq:SahuReETGwithoutdisc}
\end{equation}
for ETGs without a disc (ellipticals; rms scatter 0.60\,dex); and 
\begin{equation}
    \log_{10}(\mathrm{M}_{\mathrm{BH}} / \mathrm{M}_{\odot}) = (2.33 \pm 0.31) \log_{10}(R_\mathrm{e}) + (7.54 \pm 0.10) 
	\label{eq:SahuReLTG}
\end{equation}
for LTGs (rms scatter 0.62\,dex)

We note that \citet{Sahu_2020} derived these relations from 1-dimensional fits to high-resolution space-based data, mostly at a wavelength of 3.6\,$\mu$m (infrared). Their applicability to ground-based seeing-limited optical data of galaxies close to the resolution limit, type 1 AGN, and two-dimensional decompositions has not been tested and this is the prime interest of this paper. \\

For comparison with the \citet{Sahu_2020} relations, we also obtain the SMBH mass from our photometric measurements using the black hole mass - bulge stellar mass (M$_{*, \mathrm{bulge}}$) and black hole mass - galaxy stellar mass (M$_{*, \mathrm{tot}}$) relations presented in \citet[their equations 11 and 14]{Bentz2018}
\begin{equation}
    \log_{10}(\mathrm{M}_{\mathrm{BH}} / \mathrm{M}_{\odot}) = (1.06 \pm 0.24) \log_{10}(\frac{\mathrm{M}_{*, \mathrm{bulge}}}{10^{10} \mathrm{M}_{\odot}}) + (7.02 \pm 0.17) 
	\label{eq:Bentz2018Mbulge}
\end{equation}
with a scatter of $(0.39 \pm 0.12)$\,dex; and 
\begin{equation}
    \log_{10}(\mathrm{M}_{\mathrm{BH}} / \mathrm{M}_{\odot}) = (1.69 \pm 0.46) \log_{10}(\frac{\mathrm{M}_{*, \mathrm{tot}}}{10^{11} \mathrm{M}_{\odot}}) + (0.38 \pm 0.13) 
	\label{eq:Bentz2018Mstar}
\end{equation}
with a scatter of $(0.38 \pm 0.13)$\,dex. In both cases, we chose the relations derived with the M/L ratio of \citet{Bell2001}, since they showed lower scatter than the equivalent versions using the \citet{Into2013} M/L predictions \citep{Bentz2018}.

\citet{Bentz2018} derived these relations for a sample of low-redshift AGN with SMBH mass measurements from reverberation mapping, using a two-dimensional surface brightness decomposition in the optical $V$ and near-infrared $H$ bands. Their results should therefore be directly applicable to our work.\\

To derive the stellar masses required in equations~\ref{eq:Bentz2018Mbulge} and~\ref{eq:Bentz2018Mstar} from the $g$ and $i$ band luminosities of our surface brightness fits, we use the M/L from \citet[their equation 8]{Taylor2011}: 
\begin{equation}
    \log_{10}(\mathrm{M}_{*} / \mathrm{M}_{\odot}) = 1.15 + 0.7(g - i) - 0.4M_i 
	\label{eq:TaylorML}
\end{equation}
with an accuracy of $\sim$\,0.10\,dex, where $(g - i)$ is the rest-frame colour and $M_i$ the absolute $i$-band magnitude. Since our sample is at low redshifts, $k$-corrections are negligible; we obtain the same results by using the \citet{Bryant2015} relation based on observed-frame apparent magnitudes. Equation~\ref{eq:TaylorML} was fitted to a sample of GAMA galaxies with stellar mass estimates obtained through spectral energy distribution (SED) fits of data in the optical $ugriz$ bands. The presence of type 1 AGN in our sample may lead to deviations from this relation since the majority of GAMA galaxies - and thus the \citet{Taylor2011} sample - do not contain type 1 AGN.\\

\section{Data analysis}
\label{sec:dataanalysis}

In this section we describe the modelling of the complex spectra of type 1 AGN (Section~\ref{sec:3.1}) and the surface brightness decomposition of the host galaxy containing the studied AGN (Section~\ref{sec:sbfitting}), and how we derived SMBH masses from measured parameters using the scaling relations outlined in Section \ref{sec:scalingrelations}.

\subsection{Spectral fitting}
\label{sec:3.1}

The spectral fitting was performed on observed optical spectra of the sample retrieved from the GAMA database. Since type 1 AGN optical spectra are complex, including the contribution from the host galaxy and different components of AGN contributing to the optical emission, we describe in this section our approach to correct for those and extract the broad component of emission lines.

\subsubsection{Methodology of spectral fittings}

In order to extract the pure broad emission lines from complex optical spectra we fit the full spectral range from 4000 to 7000 \AA. Fitting a wide spectral range has been shown to be the best approach to estimate the level of underlying continuum necessary for measuring the H$\alpha$ flux and even more the line width, since the latter is sensitive to the estimated continuum level \citep[see e.g.,][]{2008MNRAS.383..581D, 2019ApJS..243...21L, 2019ApJ...886...42D}. Also, including the H$\beta$ spectral range with the prominent [O III] lines, is needed for estimating the width of narrow emission lines and later better removal of the contribution of narrow H$\alpha$ and [N II] lines from the total H$\alpha$ profile, which are often blended. Another fact in favor of full spectral range fitting is better estimating  of the contribution of Fe II complex emission, which is dependent on the Fe II detected in the blue part of the spectrum \citep[see e.g.][and references therein]{2023ApJS..267...19I}. The Fe II emission on the one hand is contributing to the H$\alpha$ line flux, but it can also mimic the Fe II quasi-continuum and thus influence the continuum level \citep[][]{2023A&A...679A..34P}. Therefore, here we decided to apply the complete spectral fitting on a wide range of wavelengths, as this case-study should present an approach to extract the broad emission line parameters,  minimizing known systematics when fitting the narrow spectral range including only the emission line of interest.

For the AGN spectral analysis, we used the python-based code for multi-component spectral fitting - \textsc{fantasy} \citep[Fully Automated pythoN Tool for AGN Spectra analYsis\footnote{Open-source code https://fantasy-agn.readthedocs.io/en/latest/}, see][for details]{2023ApJS..267...19I}, which is designed to fit AGN optical spectra in a wide range of wavelengths. Especially for the purpose of this work \textsc{fantasy} has been upgraded to include a special class of reading function \textsc{read\_gama\_fits} optimized to read spectra of the GAMA database.

\subsubsection{Preparing the spectra and host galaxy correction}

The pre-processing of the optical spectra consists of the correction for the Galactic extinction and cosmological redshift. The next part is the estimation and subtraction of the host galaxy stellar contribution. This step is especially important in the case of low-luminosity AGN for which optical spectra tend to be host-dominated, which is the case for our sample of nearby objects.
The host galaxy starlight is reconstructed through the fitting of the total observed spectra as a linear combination of galaxy and quasar eigenspectra constructed for the usage within the SDSS spectral analysis of quasars \citep{yip04a, yip04b, vanden_berk06, 2020ApJS..249...17R}.  Here we used all available eigenvectors (10 eigenvectors for galaxy (stellar) and 15 for quasar components) provided by \cite{yip04a, yip04b} from the analysis of SDSS galaxy and quasar spectra, to search for the host galaxy component. During the host fitting, we have masked strong narrow emission lines. The pure AGN component is reconstructed by subtracting the estimated host galaxy from the observed spectrum. An example of host and AGN decomposition is given in Figure \ref{fig: spectra}, top panel. The lower spectrum (solid blue line) represents the reconstructed AGN spectrum. We remind the reader that the GAMA database (i.e., its \textsc{SpecLineSFR: GaussFitComplexv05} table) contains spectral line parameters measured from the observed spectra for which the contribution of the host galaxy was not accounted for \citep[see][for details]{Gordon_2017}.

\subsubsection{Multi-component model and fittings}

Next we performed the simultaneous multi-component spectral fitting to the full range of observed AGN spectra. The main motivation of the fitting is to remove narrow and satellite lines and extract the clear broad component of the H$\beta$ and H$\alpha$ lines. In case the Fe II emission near H$\beta$ was present, we included the Fe II model that is based primarily on the atomic data of the Fe II transitions \citep[][]{kov10, 2012ApJS..202...10S, 2023ApJS..267...19I}. The model uses Gaussians of the same width and shift for every individual line profile, and groups the lines based on the same lower level in the transition. Within a single group, all lines have an intensity connected to the strongest line through the atomic data \citep{2023ApJS..267...19I}. 

We run the same fitting model for all 28 optical spectra, fitting the rest wavelength range $\sim$4000-7000 \AA. The multi-component model consisted of: 
\begin{enumerate}
    \item a broken power law continuum with a break wavelength of 5650\AA\, adopted because it avoids the wavelength regions of the prominent emission lines \citep{2023ApJS..267...19I};
    \item hydrogen H$\alpha$, H$\beta$ and H$\gamma$ lines with two broad Gaussian components, to account for the complex and asymmetric line profiles;
    \item helium broad lines He I 4471$\mathrm{\mathring{A}}$, He I 5877$\mathrm{\mathring{A}}$, and He II 4686$\mathrm{\mathring{A}}$;
    \item narrow emission lines, all fixed to have the same shifts and widths as [O III] 5007$\mathrm{\mathring{A}}$: H$\alpha$, H$\beta$ and H$\gamma$, He I 4471$\mathrm{\mathring{A}}$, He I 5877$\mathrm{\mathring{A}}$, He II 4686$\mathrm{\mathring{A}}$, [O III] 4363$\mathrm{\mathring{A}}$, [O III] 4959, 5007$\mathrm{\mathring{A}}\mathrm{\mathring{A}}$, [N II] 6548, 6583$\mathrm{\mathring{A}}\mathrm{\mathring{A}}$, [S II] 6716, 6731$\mathrm{\mathring{A}}\mathrm{\mathring{A}}$; the ratio of [O III] 4959,5007$\mathrm{\mathring{A}}\mathrm{\mathring{A}}$ and [N II] 6548,6583$\mathrm{\mathring{A}}\mathrm{\mathring{A}}$ doublets were fixed to 3 \citep[][]{2007MNRAS.374.1181D,2023AdSpR..71.1219D}; the list of used narrow lines contains also nebular line [O I] 6300$\mathrm{\mathring{A}}$ and [O I] 6364$\mathrm{\mathring{A}}$;
\item the intermediate component of the [O III] doublet to account for outflowing component often seen in AGN \citep[][]{2022A&A...659A.130K}; 
\item the optical Fe II model based on the atomic data of the transitions
\citep[see for details][] {kov10,2012ApJS..202...10S,2023ApJS..267...19I}; this component was included only for less noisy spectra (S/N ratio $\gtrsim 20$) or if clearly detected in the vicinity of H$\beta$. 
\end{enumerate}

\begin{figure}
\centering
    \includegraphics[width=\columnwidth]{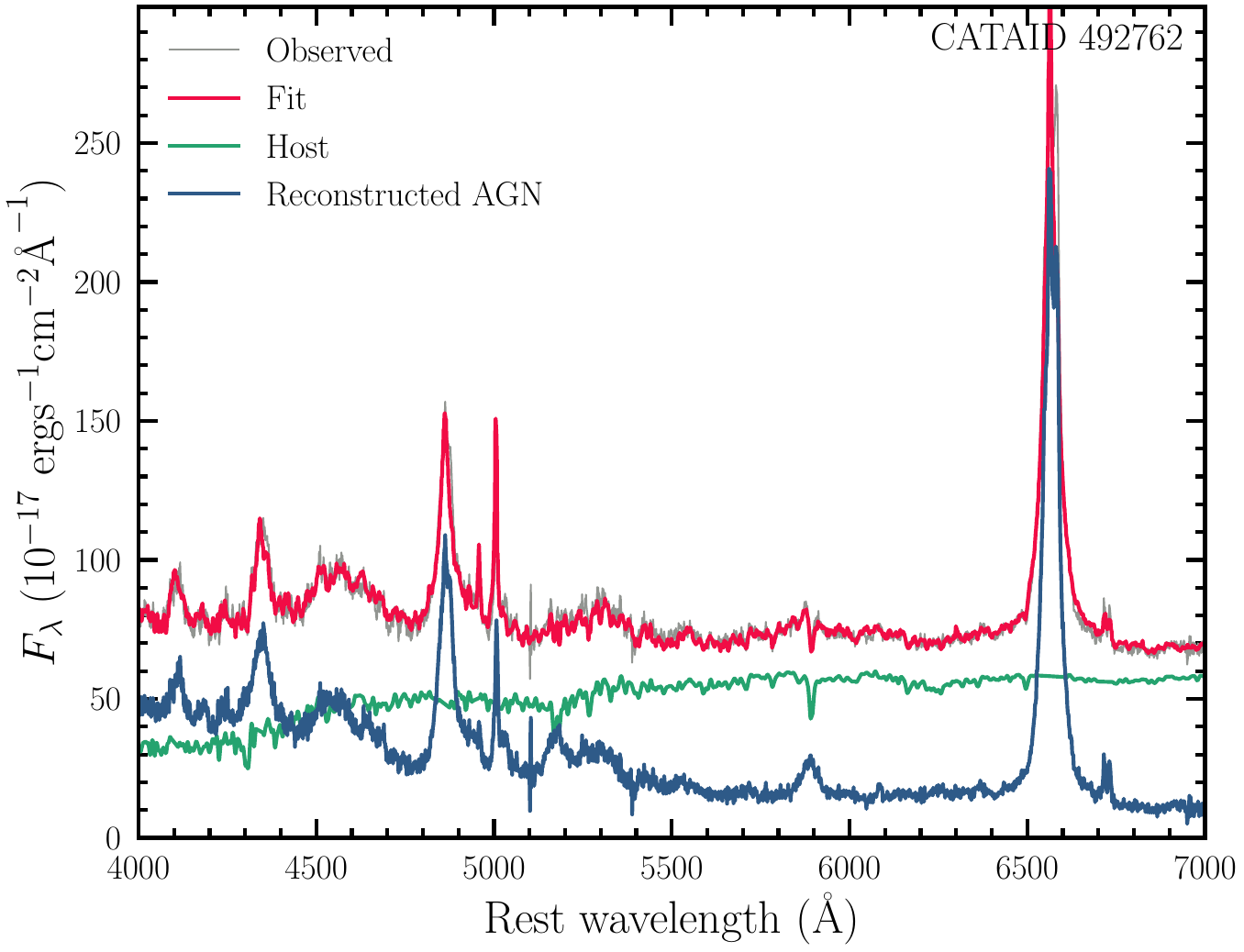}
	\includegraphics[width=\columnwidth]{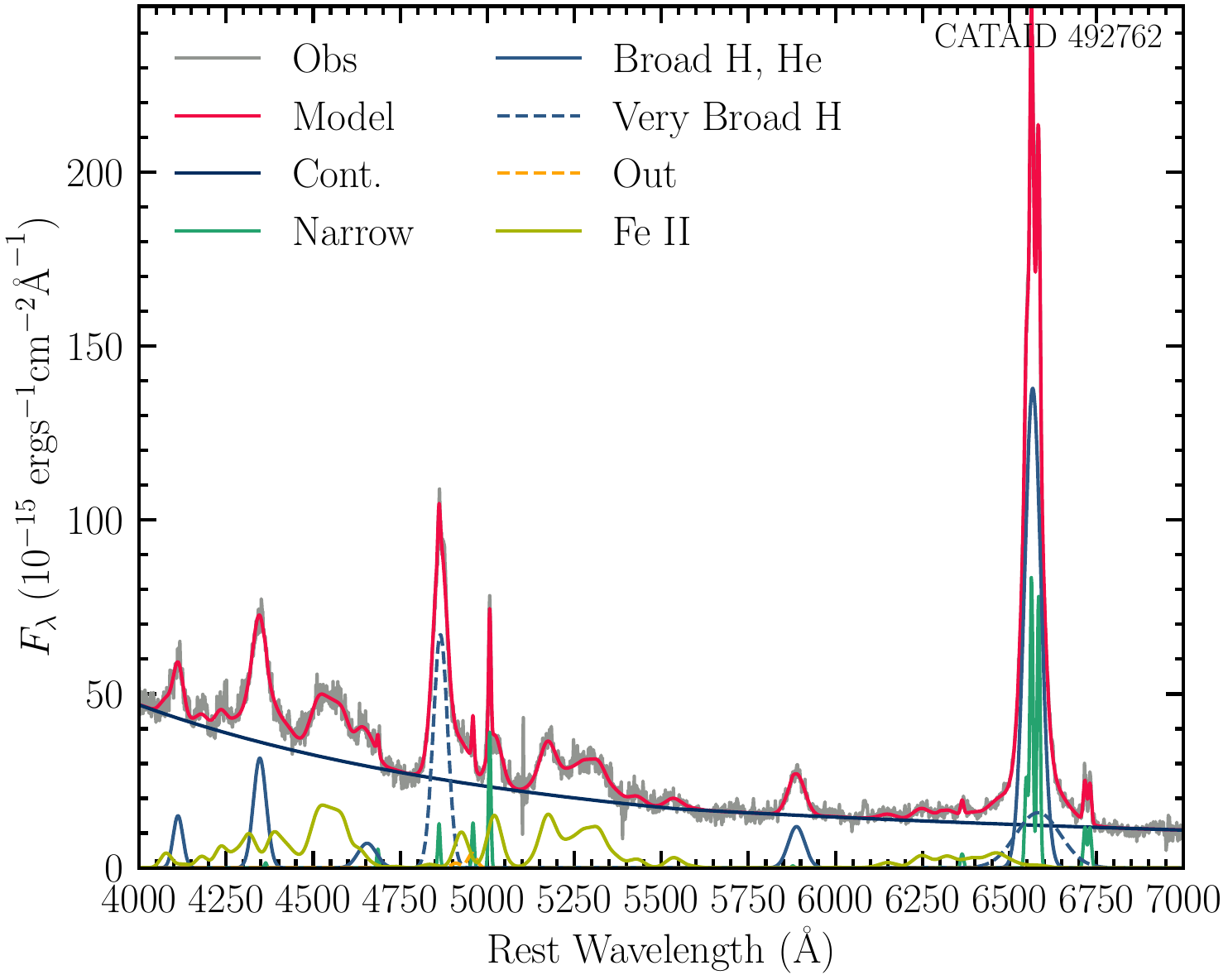}
    \includegraphics[width=0.95\columnwidth]{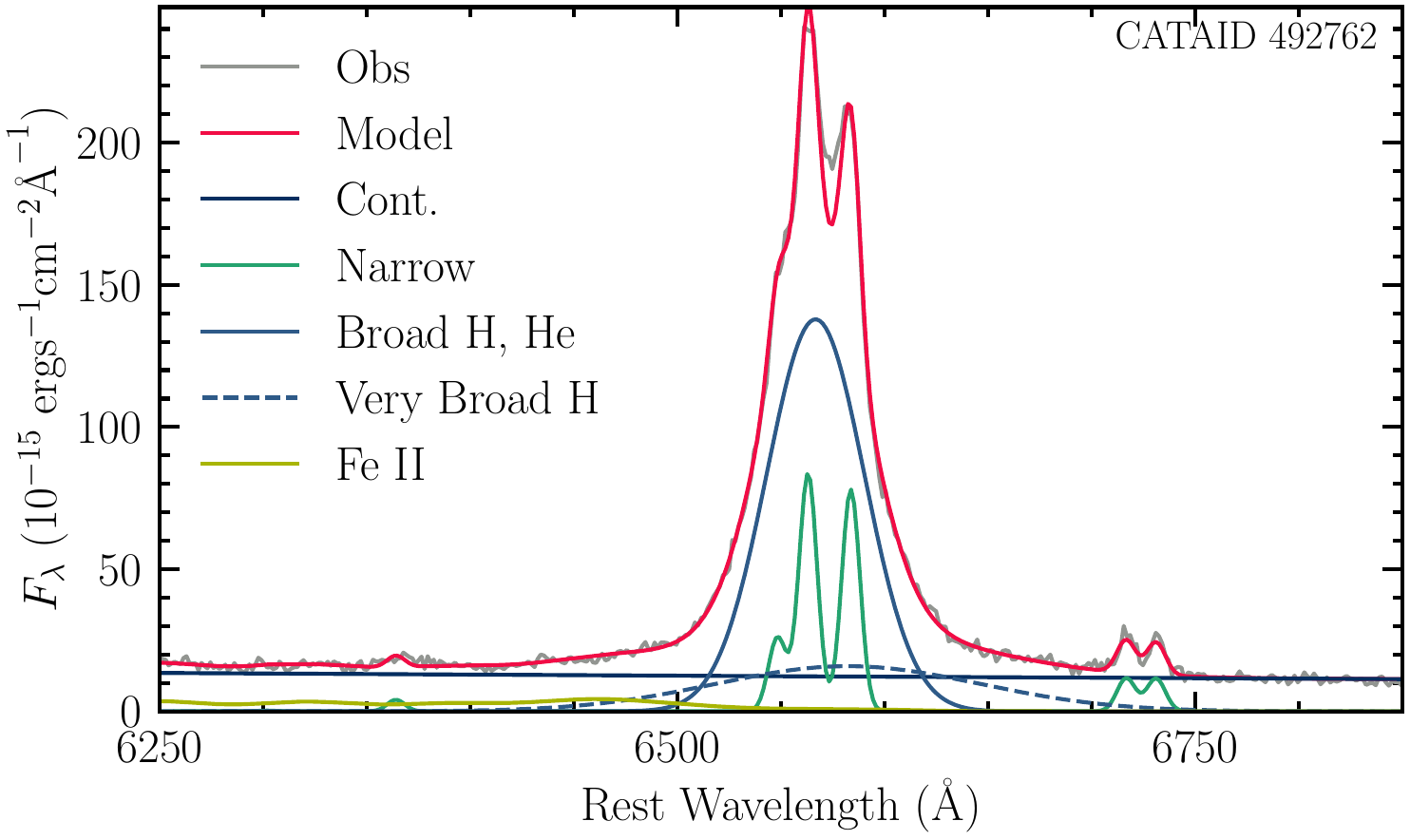}
    \caption{{\bf Top}: An example of the AGN component (blue line) recovered after subtracting the contribution of the stellar component of the host galaxy (green line) from an observed spectrum (gray line) of CATAID 492762. 
    {\bf Middle}: Multi-component fitting with the  \textsc{fantasy} code of the AGN component (gray line) in the $\lambda\lambda$4000-6800 \AA\, wavelength range. The model (red line) consists of: broken power-law (blue line) representing the underlying AGN continuum, narrow lines (green solid line), broad (blue solid line) and intermediate (blue dashed line) components of Balmer lines (H$\alpha$, H$\beta$, H$\delta$) and He I lines, intermediate (outflowing) components of [O III] lines (orange dashed line). 
    {\bf Bottom}: Zoom-in of the H$\alpha$ line region, with same notation as in the middle panel.}
    \label{fig: spectra}
\end{figure}

The total sample was fitted automatically with this model. The goodness of the best fitting results were evaluated using the reduced $\chi^2_\nu$ parameter. 
An example of the fit for an object with strong Fe II emission is given in Figure \ref{fig: spectra} (middle panel). 

\subsubsection{Measured spectral parameters and uncertainties}

We measured the line flux and the FWHM from the modelled broad H$\alpha$ line profile, which is obtained as the sum of the two broad gaussians (Figure \ref{fig: spectra}, bottom panel). 
The luminosity was calculated using the luminosity distance. 

The uncertainties of the measured quantities are estimated using the Monte Carlo approach to randomly add noise to the observed spectra and fit again with the same model \citep[see][]{2023ApJS..267...19I}. We created 50 mock spectra and calculated the uncertainties as the semi-amplitude of the range enclosing the 16th and 84th percentiles of the distribution of extracted spectral quantities \citep[as in][]{2020ApJS..249...17R}. These are reflecting only statistical errors, and are probably underestimated with respect to the true uncertainties. Potential sources of systematics could be a model choice, such as the selection of the Fe II model, the estimate of the underlying continuum, and the subtraction of the narrow and satellite lines \citep[e.g. see discussions in][]{Greene05, Xiao_2011, 2012ApJ...755..167D}. 
In further analysis we use the above somewhat lower uncertainties of the spectral parameters, which are reflected in the spectral mass uncertainties, as those have not been used to derive any relation. 
We note that omitting the Fe II component significantly increased the $\chi^2_\nu$.

We then obtain a SMBH mass from the H$\alpha$ width and luminosites (obtained from the total broad line) using Equation~\ref{eqn:xiao}, with mass uncertainties propagated from the uncertainties of H$\alpha$ line width and luminosity. The SMBH mass calculated using Equation \ref{eqn:xiao} are in the expected range of $10^6-10^8\,\text{M}_\odot$ for type 1 AGN.\\

\subsection{Surface brightness fitting}
\label{sec:sbfitting}
The surface brightness fitting was performed on i-band and $g$-band imaging data from the KiDS data release 4 \citep[][]{Kuijken2019} using the Bayesian two-dimensional profile fitting code \textsc{ProFit} \citep{Robotham2017} and treating both bands independently from each other. In a first step, the galaxies are run through the automated bulge-disc decomposition pipeline described in \citet[][]{Casura2022}, which is the basis for the \textsc{BDDecomp} DMU on the GAMA database. As a result, we obtain postage stamp cutouts of the galaxies of interest in the $i$ and $g$ bands, and associated sigma (error) maps, masks, segmentation maps and sky (background) maps; PSF estimates with corresponding diagnostic plots; and three model fits to each galaxy also with diagnostic plots and ancillary information.
We use the automated single S\'ersic fits to obtain overall galaxy colours; and the remaining data returned by the pipeline as inputs for the multi-component surface brightness fitting performed in this work, which we describe in this section.

\subsubsection{Visual galaxy classification and model definition}
\label{sec:visualgalaxyclassification}
We determine the model to be fitted to each galaxy based on its visual morphological classification. We aim to distinguish between elliptical (E) galaxies, lenticular (S0) galaxies, spiral (S) galaxies and barred spiral (SB) galaxies; where the former two belong to the class of ETG and the latter two are LTG.

The GAMA database provides several catalogues of visual classifications for different samples of galaxies and with different classification schemes, which are collected in the \textsc{VisualMorphologyv03} DMU: a classification as "Elliptical" or "not Elliptical" following \citet{Driver2012}, a Hubble type classification based on six expert opinions following \citet{Kelvin2014}, a probability that a galaxy is disturbed presented by \citet{Robotham2014} and a Galaxy Zoo classification giving the de-biased fraction of voters that considered a certain galaxy to be elliptical or not, with the corresponding analysis presented in \citet{Lintott2011}. In addition, the \textsc{gvkMorphologyv02} DMU assigns a class to $z<0.08$ galaxies in GAMA DR4.0, see \citet{Driver2022}.

20 of our sample of 28 galaxies were classified in at least one of those catalogues. For the remaining eight objects, and for the three cases where the classification between different GAMA catalogues disagreed, we used our own visual inspection of colour images, leaving only one ambiguous case (\textsc{CATAID} 506001, eventually classified as S0). Galaxies that showed signs of disturbance were flagged to be treated with additional care. 

We show the final classification for each object, along with the model fitted, in Table~\ref{tab:sbfittingresults}. In total, we have 10 elliptical galaxies modelled with a single S\'ersic component, 7 lenticulars and 5 spirals, both modelled with a S\'ersic bulge plus exponential disc, and 6 barred spiral galaxies with an additional Ferrers profile for the bar component. Two objects (one S0 and one SB) required broken exponential discs. In addition, all objects have an AGN component modelled as a point source since all galaxies of our sample have a visible type 1 AGN at their centre by construction (cf. Section~\ref{sec:datasample}). Where the visual inspection (Section~\ref{sec:segmentationmaps}) indicated the presence of a foreground star or nearby galaxy, we also model those; as a point source or single S\'ersic object respectively.

\subsubsection{Segmentation maps and nearby objects}
\label{sec:segmentationmaps}
The segmentation maps indicate the pixels that should be considered during the fit (this is equivalent to setting the weight of all pixels outside the segment to zero during the fit). For the \textsc{BDDecomp} DMU they are obtained in an automated fashion using the source extraction and image analysis package \textsc{ProFound} \citep{Robotham2018}, see \citet{Casura2022} for details of the procedure. For the 28 galaxies considered in this project, we visually inspected all segmentation maps and optimised them manually as appropriate. Secondary objects that overlap with the object of interest were added to the main galaxy segment along with initial guesses for its position in order to allow for a joint fit (cf. Section~\ref{sec:visualgalaxyclassification}). Objects with nearby saturated stars that critically affect the data quality were flagged (see Table~\ref{tab:sbfittingresults}).

\subsubsection{PSFs}
\label{sec:psfs}
Accurate PSFs are particularly important for the present study since the bulge parameters critically depend on it, especially due to the presence of an unresolved but bright (potentially dominant) AGN which is modelled with a point source; meaning that it has the PSF profile shape with no additional degrees of freedom. Relevant discussions can be found in, e.g., \citet{Bentz2018, Zhao2021}, and the detailed description of the systematic uncertainties associated with incorrect PSF estimates in the context of AGN plus host image decomposition in \citet{Zhuang2023}. Since the automated model PSFs described in \citet{Casura2022} start to show deviations from the true PSF for very bright point sources \citep[cf.][]{mythesis}, this resulted in residuals appearing at the centres of several of our galaxies, indicative of a PSF model inadequacy.

We therefore re-fitted the PSFs for all of our 28 galaxies using a double Moffat function and jointly fitting all suitable stars around the galaxy of interest with a full Markov-chain Monte Carlo (MCMC) treatment. The suitable stars are selected from the candidate list of objects previously fitted with single Moffat functions, but applying more generous criteria relative to the procedure described in appendix C of \citet{Casura2022}. In addition, we significantly expand the star segment sizes in order to include more of the PSF wings in the fit. These choices lead to model PSFs that are based on a minimum of three and a maximum of 24 stars fitted jointly, with most galaxies having 10 - 20 suitable stars. The segments used for fitting include between 94\,\% and 99\,\% of the total model flux obtained by extrapolating the profile to infinity.

The resulting model PSFs show strong improvements over the single Moffat versions, in particular for bright stars, 
although in some cases, asymmetries or variations in the residuals between individual stars remain. These objects are considered to have a ``poor PSF'' and are marked in Table~\ref{tab:sbfittingresults}. We comment on the effect of the PSF choice on the final outcome and consider systematic uncertainties introduced by the PSFs in Section~\ref{sec:errorestimation}.
We note that the procedure applied in this study is computationally prohibitive for large samples, with single-CPU run-times for PSF estimations for a single galaxy ranging between $\sim$$1$ and $10$ days; as opposed to a few minutes for the routine described in \citet{Casura2022}.

\subsubsection{Free parameters and constraints}
\label{sec:fitconstraints}

The AGN component (point source) has three free parameters: position ($x$ and $y$; tied together for all components as recommended by \citealt{Zhuang2023}) and total magnitude $m$. We constrain the AGN magnitude to be at least as bright as the the intrinsic (non-PSF-convolved) spheroid flux within one FWHM of the PSF to ensure a notable AGN contribution and suppress unphysically high spheroid S\'ersic indices. Table~\ref{tab:sbfittingresults} indicates which fits were affected by this.

The spheroid (S\'ersic profile) has five free parameters: the magnitude $m$, effective radius $R_\mathrm{e}$, S\'ersic index $n$, axial ratio (minor/major axis) $b/a$ and the position angle PA. Following \citet{Casura2022}, we set the \textsc{boxyness} to 0. 

Exponential discs have four parameters since an exponential is a S\'ersic function with $n\,=\,1$. Broken exponential discs have three additional parameters: an inner and an outer scale length $h_1$ and $h_2$ instead of $R_\mathrm{e}$, the break radius $R_\mathrm{b}$ at which the profile changes from $h_1$ to $h_2$, and a parameter $a$ describing the sharpness of this transition, which we allow to vary between 0.1 and 1 \citep[following][]{Erwin2008}. 

Bars (Ferrers profile) are fitted with five free parameters: $m$, $b/a$, PA, a truncation radius $R_\mathrm{out}$ beyond which the flux is set to zero and a shape parameter, $\alpha$, which controls the overall profile slope and the sharpness of the truncation as we approach $R_\mathrm{out}$. The \textsc{boxyness} is set to 1 to reflect the boxy shape of typical bars; and the second shape parameter $\beta\,=\,0$. This follows the recommendation of \citet{Ciambur2016} and is equivalent to using an original (non-modified) Ferrers profile \citep{Kim2015}. It differs from the choice of \citet{Blazquez-Calero2020}, who fixed $\alpha$ instead, although they discuss that fixing $\beta$ would also be a viable alternative. In our case, we fixed $\beta$ to avoid the bar profile becoming cuspy in the central region. For a visual impression of the effects of $\alpha$ and $\beta$ on the profile shape, see figure~4 in \citet{Peng2010} and figure~6 in \citet{Ciambur2016}. 

Overlapping stars (point sources) or galaxies (S\'ersic profiles) have 3 and 7 parameters, respectively, including position in $x$ and $y$, which is allowed to vary within a region of five pixels (1\,arcsec) around the manually defined initial guess (cf. Section~\ref{sec:segmentationmaps}). No significant degeneracies with the parameters of the main object occur since they are spatially separated. 

\begin{figure}
\centering
    \includegraphics[width=\columnwidth]{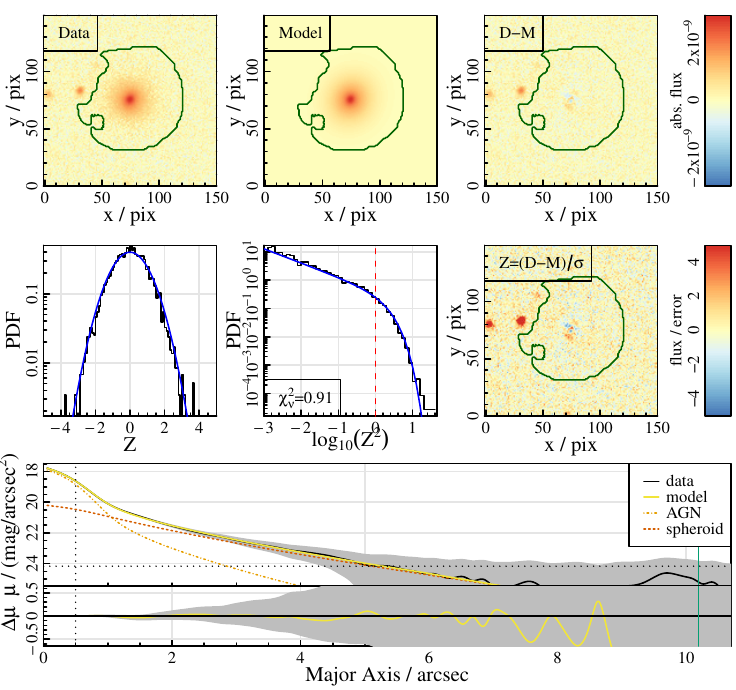}
    \caption{The surface brightness fit to galaxy 218688, classified as elliptical (8 free fitting parameters). The \textbf{top row} shows the data, model and residual (data - model) in absolute values of flux given by the colour bar on the right (non-linear scale); the green contour indicates the segment used for fitting. The \textbf{middle row} shows goodness of fit statistics, namely the distribution of normalised residuals $Z$~=~(data - model)/error relative to a Gaussian; the PDF of $Z^2$ compared to a $\chi^2$-distribution with one degree of freedom; and the two-dimensional distribution of $Z$ capped at $\pm\,5\sigma$ (colour bar on the right). The \textbf{bottom row} shows the model and its individual components compared against the data in one-dimensional form (azimuthally averaged over elliptical annuli). The FWHM of the PSF and the approximate $1\,\sigma$ surface brightness limits are indicated by vertical and horizontal dotted lines for orientation; the vertical solid green line indicates the segment radius. The pixel scale is 0.2\,arcsec for KiDS data, i.e. 1\,arcsec corresponds to 5\,pix.}
    \label{fig:examplefitE} 
\end{figure}

\begin{figure}
\centering
    \includegraphics[width=\columnwidth]{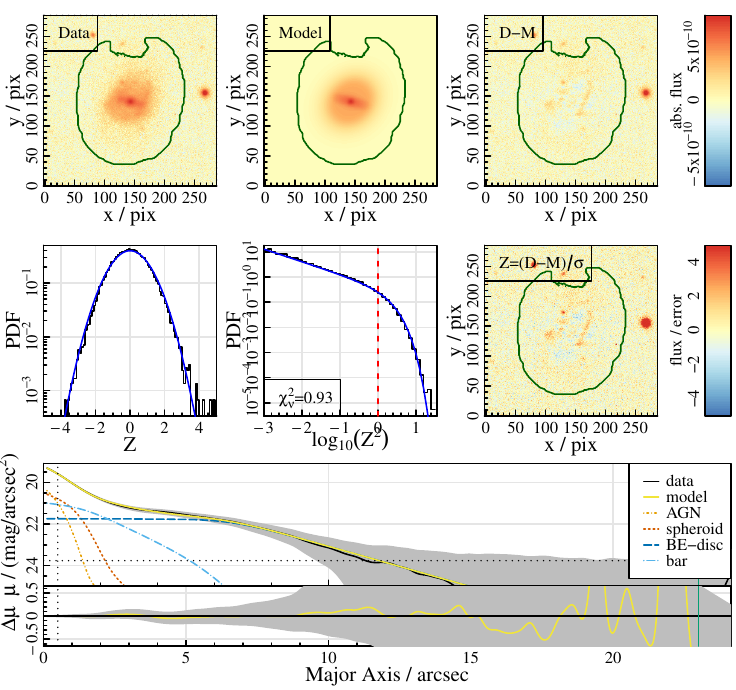}
    \caption{The $i$-band surface brightness fit to galaxy 3631902, classified as a barred spiral galaxy with a broken exponential disc (20 free fitting parameters). Panels are the same as in Figure~\ref{fig:examplefitE}.}
    \label{fig:examplefitSB} 
\end{figure}

The total number of free fitting parameters for each object is listed in Table~\ref{tab:sbfittingresults}; and Figures~\ref{fig:examplefitE} and~\ref{fig:examplefitSB} show examples of galaxies fitted with a low and high number of fitting parameters, respectively (corresponding corner plots in Figures~\ref{fig:cornerplotE} and~\ref{fig:cornerplotSB}). Finally, we note that the scale parameters $n$, $b/a$, $R_\mathrm{e}$, $R_\mathrm{b}$, $R_\mathrm{out}$, $h_1$, $h_2$, $a$ and $\alpha$ are treated in logarithmic space throughout, i.e. the actually fitted parameter for the S\'ersic index is $\log_{10}(n)$ and likewise for all others.

\subsubsection{Initial guesses and swapped components}
\label{sec:initialguesses}
Bulge-disc swapping (meaning the S\'ersic function fits the disc and the exponential fits the bulge) is a common problem in galaxy fitting, see, e.g., \citet{Allen2006, Head2014, Mendel2014, Meert2015, Lange2016, Cook2019, Barsanti2021, Casura2022}. It is particularly prominent for multi-component fits with (nearly) interchangeable model components. With up to 4 components, particularly the LTG in our sample showed frequent bulge-disc, bulge-bar, or bulge-disc-bar swapping. This can in theory be alleviated by imposing appropriate fitting constraints or choosing informative priors. In practice, it is often sufficient (and more convenient) to define initial guesses close to the desired likelihood maximum, exploiting the fact that even a full MCMC routine will show a (mild) dependence on the starting values for finite run-times \citep{Head2014, Lange2016, Cook2019, Barsanti2021, Casura2022}. 

We choose the latter approach and provide initial values for each $i$-band fit, which are manually optimised to - and successful in - avoiding swapped components. Details on how these are derived are given in Appendix~\ref{app:initialguesses}.

\subsubsection{Fit and convergence}
\label{sec:convergence}
We now have all input data for the final galaxy (re-)fits: the cutouts from KiDS $i$-band and $g$-band images, sky background estimates and sigma maps from the \textsc{BDDecomp} DMU pipeline (Section~\ref{sec:datasample}; separately for each band), the improved segmentation maps including nearby objects for joint fitting (Section~\ref{sec:segmentationmaps}; the same for both bands), the improved double-Moffat jointly fitted PSFs (Section~\ref{sec:psfs}; separate in each band), the visual morphological classification of each galaxy and correspondingly the model to be fitted (Section~\ref{sec:visualgalaxyclassification}; the same in both bands), the parameters to be freely fitted, fixed or constrained for each component (Section~\ref{sec:fitconstraints}; also the same in both bands), and reasonable initial guesses for all parameters of the main galaxy and jointly fitted sources (Section~\ref{sec:initialguesses}; the same for both bands, originally derived for the $i$-band). As in \citet{Casura2022}, we obtain the best-fitting parameters independently in both bands using \textsc{ProFit} with a Normal likelihood function in combination with the \textsc{convergeFit} function from the \textsc{AllStarFits} package \citep{AllStarFit}. This function combines a number of downhill gradient optimisers with several MCMC algorithms to characterise the many-dimensional shape of the likelihood maximum in a robust, fully automated and computationally affordable way.

In detail, \textsc{convergeFit} runs several downhill gradient algorithms available in the \textsc{nloptr} package \citep{nloptr}. After these have converged, an MCMC chain with 500 iterations using the Hit-and-Run-Metropolis (HARM)\footnote{Both HARM and CHARM are variants of the Hit-And-Run algorithm introduced by \citet{Turchin1971}.} algorithm in the \textsc{LaplacesDemon} package \citep{LaplacesDemon} is started. This process is repeated until the fractional change in the log-likelihood between two consecutive chains (batches of 500 samples) is less than $e$, which is the stopping criterion chosen by \citet{Taranu2017}. Finally, the Componentwise Hit-and-Run-Metropolis (CHARM), also from the \textsc{LaplacesDemon} package, is run with 2000 iterations and again repeated until the stopping criterion is reached (usually only once since the chains have converged already using HARM). CHARM is more robust but also more computationally expensive than HARM, hence it is only called in the end to collect likelihood samples around the peak which has already been found in previous steps. As shown in \citet{mythesis}, the 2000 final likelihood samples returned by CHARM can be considered stationary for all practical purposes; and thus they are used in further analysis of the galaxy.

Total single-CPU run-times for our galaxies range between a few minutes for some of the simplest galaxies to several hours for those with many parameters. Example fits are shown in Figures~\ref{fig:examplefitE} and~\ref{fig:examplefitSB}, with the corresponding corner plots in Figures~\ref{fig:cornerplotE} and~\ref{fig:cornerplotSB}.

As a final test, we re-ran all galaxies on their final settings once more, except that we do not include a spheroid component in the model. This results in much longer run-times and significantly worse fits for 19 galaxies (mean increase of a factor of 9 in $\chi^2_\nu$). The remaining 9 galaxies show no or only a slight degradation of the fit quality, with $\chi^2_\nu$ increasing by less than 25\,\% (in 6 cases $\lesssim$\,10\,\%). We flag these galaxies for further inspection (see Section~\ref{sec:sbfittingresults}).

\subsubsection{SMBH mass calculation and uncertainties}
\label{sec:errorestimation}

We calculate four different photometric SMBH mass estimates: 
\begin{enumerate}
    \item M$_{\mathrm{BH}}$(spheroid S\'ersic index), 
    derived from the spheroid S\'ersic index with equations~\ref{eq:SahurelationETG} and~\ref{eq:SahurelationLTG} 
    (panel (a) in Figure~\ref{fig:mass});
    \item M$_{\mathrm{BH}}$(spheroid effective radius), 
    derived from the spheroid effective radius with equations~\ref{eq:SahuReETGwithdisc}, \ref{eq:SahuReETGwithoutdisc} and~\ref{eq:SahuReLTG} 
    (panel (b) in Figure~\ref{fig:mass});
    \item M$_{\mathrm{BH}}$(total stellar mass),
    derived from the total stellar mass of the galaxy with equation~\ref{eq:Bentz2018Mstar}, 
    where the stellar mass is obtained from our $g$ and $i$ band single S\'ersic fits with equation~\ref{eq:TaylorML}
    (panel (c) in Figure~\ref{fig:mass});
    \item M$_{\mathrm{BH}}$(spheroid stellar mass),
    derived from the spheroid stellar mass with equation~\ref{eq:Bentz2018Mbulge}, 
    where the spheroid mass is obtained from our $g$ and $i$ multi-component decompositions with equation~\ref{eq:TaylorML}
    (panel (d) in Figure~\ref{fig:mass}).
\end{enumerate}

Fitting uncertainties for all S\'ersic parameters are obtained from the corresponding posterior distributions using the $1\sigma$-quantiles as lower and upper limits, and converting to linear space for scale parameters. To account for systematic effects, we then increase the uncertainty by the ``error underestimate'' factor derived in \citet[][their table~6]{Casura2022}. These factors (how much the random error is underestimated relative to the total error including systematics) were derived from a detailed investigation of the difference between single S\'ersic $r$-band fits to the same galaxy in different exposures. We expect $g$ and $i$ band multi-component fits to be affected by similar systematics and hence apply the same corrections here. To account for PSF model inadequacies, which are not included in the error underestimate factors, we additionally increase the errors by a factor of 2. This factor was derived from the median difference between the S\'ersic index fitted to a galaxy with the single Moffat or double Moffat PSF versions, normalised by its error (including the systematics correction). 

The resulting uncertainties are typically symmetric and amount to approximately 0.4\,mag in spheroid magnitude (for both the $g$ and $i$ bands), $\sim$\,3\,\% in effective radius and 14\,\% in S\'ersic index.
In deriving these uncertainties, we do not take parameter correlations into account. As can be seen in Figure~\ref{fig:cornerplotE} \citep[see also the discussion in][]{Casura2022}, the spheroid S\'ersic index, effective radius and magnitude as well as the point source magnitude are typically correlated with each other; with many more degeneracies arising in fits with higher numbers of components (Figure~\ref{fig:cornerplotSB}). Partially, these are accounted for by the error underestimate factors \citep[see][]{Casura2022}, but especially for multi-component fits, parameter correlations will introduce an additional error budget. Thus, even with the systematics corrections, our derived uncertainties are likely underestimates.

Uncertainties on the SMBH mass estimates are derived from the uncertainties of the relevant S\'ersic parameters and the uncertainties of the used scaling relations where provided; using standard error propagation assuming uncorrelated errors.

\section{Results and discussion}
\label{sec:results}

Here we describe the results of the spectral modelling (Section~\ref{sec:spectralfittings}) and the surface brightness modelling (Section~\ref{sec:sbfittingresults}) before comparing the respective SMBH mass estimates against each other (Section~\ref{sec:masses}).

\subsection{Spectral fitting}
\label{sec:spectralfittings}

Table \ref{tab:sbfittingresults}, columns 5-7, list the measured spectral parameters of the total broad H$\alpha$ line, namely the luminosity and the FWHM, extracted from the multi-component fits as a sum of two broad components; and the SMBH mass calculated from them using equation~\ref{eqn:xiao}, M$_{\mathrm{BH}}$(H$\alpha$). Figure \ref{fig: spectra} shows an example fit for object CATAID 492762.

The total sample of 28 type 1 AGN was fitted automatically with the same model consisting of the underlying continuum, narrow lines, and broad emission lines, yielding reasonable results for the majority of objects. The complex Fe II model was included in case of better quality spectra and if clearly detected near H$\beta$ \citep{2023ApJS..267...19I}. We note that for two objects, CATAIDs 362697 and 609069, the previous detection of the broad H$\alpha$ and H$\beta$ lines reported in the GAMA database (table GaussFitComplexv05) was probably the result of the noise fitting. Our analysis revealed that there is no broad component.
The model captured a broad artifact in the range of the H$\alpha$ and [N II] lines, which is of two orders of magnitude lower luminosity than the average value in the sample. This is either noise or contribution from much weaker narrow components coming from kinematically different emitting line regions, which can often be hidden or mimic the broad component \citep[][]{2022A&A...659A.130K}. These two objects remain in the sample to illustrate the source of potential uncertainty and false detection of the broad component in automated spectral fittings, especially if using a single multi-component model while treating a large data sample.

Despite the low S/N ratio of the observed spectra, which was in some cases below 5, the \textsc{fantasy} code was able to give reasonable fit results in all cases. All of our AGN are located in nearby extended galaxies, thus in most cases the optical spectra are host dominated with clearly visible strong absorption lines (top panel in Figure~\ref{fig: spectra}). The biggest challenge in the spectral fitting was the extraction of the host stellar contribution, which in some cases led to issues in the location and shape of the underlying AGN continuum (middle panel in Figure~\ref{fig: spectra}). Note that in the case of spectra taken by the GAMA team, there is an additional uncertainty in the continuum level due to the instrument setup which required that the blue and red part of the spectra are spliced together
\citep{Baldry_2014}. Nevertheless, this had no strong influence on the extraction of the H$\alpha$ broad line and further measurements of its spectral properties.

In our sample of type 1 AGN we have representatives of different subtypes in terms of their spectral properties (e.g., emission line width, strength of Fe II), and consequently physical conditions. 
Few cases are identified as narrow-Seyfert 1 (NLSy1) objects showing narrower broad lines and very strong iron emission \citep{2017ApJS..229...39R}. It was shown that Fe II emission may contaminate the H$\alpha$ line wings, and consequently the measurements of the widths and luminosities \citep[see e.g.][]{2023ApJS..267...19I}, which may become important in case of NLSy1 objects. Even though this contribution is small, within our analysis, this has been accounted for as the Fe II model included the wavelength range around the H$\alpha$ line (Figure \ref{fig: spectra}). 

In summary, we conclude that the spectral fits yielded reasonable results for all objects based on the assessments of $\chi^2_\nu$, with the broad H$\alpha$ component detected in 26 out of 28 objects. We tested how different fitting models would effect the $\chi^2_\nu$. For example, in case of object when Fe II emission is seen near H$\beta$ line if the fitting model was not including the Fe II component in the vicinity of the H$\alpha$ blue wing, the $\chi^2_\nu$ would be significantly increased.

Since the procedure for  decomposition of AGN optical spectra to extract the broad emission line parameters contains well established steps and has been tested for large samples of data \citep[see e.g.,][]{2011ApJS..194...45S, 2017ApJS..229...39R, 2020ApJS..249...17R, 2023ApJS..267...19I}, we focus in much greater detail on the assessment of the extraction of SMBH masses from the photometric data.

\subsection{Surface brightness fitting}
\label{sec:sbfittingresults}

Table~\ref{tab:sbfittingresults}, columns 8 - 13, list the main results of the surface brightness fitting, including the fitted model and total number of free parameters, the spheroid S\'ersic index with its uncertainty, the corresponding SMBH mass M$_{\mathrm{BH}}$(spheroid S\'ersic index), and comments on the fit quality.  We focus on the $i$-band spheroid S\'ersic index here, since the \citet{Sahu_2020} $n$ - M$_{BH}$ relation is applied to type~1 AGN in ground-based optical imaging for the first time; and thus it is of particular interest to test the reliability of this method. All other SMBH mass estimates are listed in Table~\ref{tab:masses}.

A fit is considered to have ``failed" if 
\begin{enumerate}
    \item the galaxy falls into a masked region of data (we use the masks provided by KiDS; this affects two objects near saturated stars);
    \item the uncertainty range of the (spheroid) S\'ersic index before systematics correction included either of its fitting limits (i.e. the fit hit the limit; this affects four fits in the $i$-band and five in the $g$-band multi-component decompositions, none in the single S\'ersic fits);
    \item the automated fit did not return a result (meaning the fit attempt failed and threw an error for one $i$-band single S\'ersic fit)
    \item the spheroid component was swapped with any other component in the multi-component fits, meaning that the S\'ersic profile fitted the bar or disc instead of the spheroid (this affects six multi-component $g$-band fits and zero $i$-band fits due to the way in which we constructed initial guesses, cf. Appendix~\ref{app:initialguesses}). 
\end{enumerate} 
In total, this leads to three failed single S\'ersic fits (2 masked, 1 crashed; white markers in panel (c) of Figure~\ref{fig:mass}), six failed $i$-band multi-component fits (2 masked, 4 hit limits; white in panels (a) and (b) of Figure~\ref{fig:mass}, and see Table~\ref{tab:sbfittingresults}), and 12 fits failed in either or both of the $g$ and $i$-band multi-component fits (white in panel (d) of Figure~\ref{fig:mass}: 2 masked, 5 hit limits in $g$ - of which 3 also hit their limits in $i$; and 6 swapped in $g$ - of which one is masked and one hit the limit in $i$). All fits converged according to our stopping criterion (Section~\ref{sec:convergence}). 

Table~\ref{tab:sbfittingresults} shows that fits with lower numbers of fitting parameters generally give more reliable results with smaller uncertainty ranges than those with more components. Barred spiral galaxies show a high failure rate, and many of them can be equally well fitted without the spheroid component. This indicates that the data quality, especially its resolution, is not high enough to meaningfully constrain bar components in addition to the AGN and spheroid, with 17 parameters to be determined from the central few resolution elements. Neglecting bars in the fit, however, results in the spheroid component fitting the bar, making it unsuitable for SMBH mass estimation. 

A similar issue is observed for objects with a broken exponential disc: in both cases, two of their disc parameters hit their upper limit ($a$ and $h_1$, cf. Section~\ref{sec:fitconstraints}), indicating that there are too many degrees of freedom in the fit; while fitting exponential discs instead results in a clear degradation of the fit quality.

Unbarred spiral and lenticular galaxies have fewer fitting parameters (12), resulting in higher confidence fits in general, although with exceptions. One potential issue is that galaxies may host central components that do not correspond to classical spheroids (in addition to or instead of the latter), e.g. pseudo-bulges or nuclear lenses, which we cannot distinguish at the resolution of our data. We suspect that some of our spheroid fits are affected by such additional or alternative sources of light, especially those with spheroid S\'ersic indices $n\,\lesssim\,1$ and/or elongated shapes (see comments in Table~\ref{tab:sbfittingresults} and Appendix~\ref{app:discusssahu}). 

For elliptical galaxies, the degrees of freedom are drastically reduced (8 free parameters), resulting in more stable fits. The main concern in this case are poor PSFs, leading to an imperfect subtraction of the AGN contribution and clear residuals at the galaxy centre. This is not the case for the spiral galaxies with poor PSFs, very likely because their fits have more degrees of freedom with which they can compensate for the imperfect PSF estimate. We consider this as an additional systematic uncertainty (cf. Section~\ref{sec:errorestimation}), although the effect on the parameters of an individual galaxy may be much larger than this average value. 

We also believe PSF mismatches to be a reason for AGN hitting their magnitude constraint (Section~\ref{sec:fitconstraints}), which mostly affects ellipticals, and is frequently associated with poor PSFs (Table~\ref{tab:sbfittingresults}). Without the constraint, eight galaxies had a negligible AGN flux contribution, with the spheroid accounting for all of the central flux through a high S\'ersic index. This shows that the AGN component is degenerate with the spheroid S\'ersic index to some extent (see also discussion in e.g., \citealt{bentz09}), especially for compact spheroids and imperfect PSFs. The constraint successfully suppresses unphysically high S\'ersic indices, but also introduces a non-trivial correlation between the AGN magnitude and the parameters of the spheroidal component. For these objects, therefore, the choice of constraint (minimum AGN brightness) poses an additional systematic uncertainty.

\begin{landscape}
\begin{table}
	\centering
	\caption{Overview of the spectral and $i$-band surface brightness fits of 28 objects listing: GAMA \textsc{CATAID}, the galaxy type from visual classification (E~=~elliptical, S0~=~lenticular, both ETGs; S~=~spiral and SB~=~barred spiral, both LTGs), redshift, broad H$\alpha$ luminosity and FWHM, SMBH mass calculated from the H$\alpha$ line, the components fitted to the main object (S~=~spheroid, D~=~(exponential) disc, BE-D~=~broken exponential disc, B~=~bar) plus any nearby objects (star and/or galaxy), and the total number of free fitting parameters (where we separate those of the main galaxy and nearby objects with a `+'), the spheroid S\'ersic index $n$ and the SMBH estimated from it, and comments highlighting some main uncertainties in the $i$-band surface brightness fit. SMBH masses calculated from all other methods are listed in Table~\ref{tab:masses}.}
	\label{tab:sbfittingresults}
	\begin{tabularx}{\linewidth}{rrcccccllccX}
    & & & & & & log$_{10}$( & & & & log$_{10}$( & \\
 	ID & \textsc{CATAID} & type & z & L(H$\alpha$) &w(H$\alpha$) & M$_{\mathrm{BH}}$(H$\alpha$)) &  fitted components  & N$_{\rm param}$ & $n$ & M$_{\mathrm{BH}}$(spheroid S\'ersic index)) & fit quality comments\\ 
    &    &  &  &  [10$^{41}$ erg/s] &  [km/s]  &   [M$_{\odot}$]  &  & &  & [M$_{\odot}$] & \\
    \hline
      1  &  23104   &  S0 & $0.078$ &  3.530 $\pm$ 0.027  & 5029$\pm$35  & 7.64$\pm$0.09  & AGN + S + D + star & 12 + 3   &  $0.50_{-0.07}^{+0.07}$ &  5.07$_{-0.37}^{+0.37}$ & spheroid elongated \\
     2  &  47558   &  SB & $0.070$ &  1.260 $\pm$ 0.015  & 5995$\pm$103 & 7.60$\pm$0.10  & AGN + S + D + B + star & 17 + 3   &  $0.10_{-0.02}^{+0.02}$ &  3.75$_{-0.55}^{+0.55}$ & FAILED: $n$ hit lower limit, AGN hit constraint\\
     3  &  323124  &  S  & $0.050$ &  0.330 $\pm$ 0.012  & 3520$\pm$173 & 6.83$\pm$0.12  & AGN + S + D         & 12       &  $3.79_{-0.26}^{+0.27}$ &  8.19$_{-0.17}^{+0.17}$  & FAILED: galaxy masked entirely, saturated star very nearby \\
     4  &  278841  &  S0 & $0.052$ &  2.120 $\pm$ 0.015  & 2899$\pm$35  & 7.09$\pm$0.09  & AGN + S + BE-D     & 15       &  $0.99_{-0.13}^{+0.15}$ &  6.24$_{-0.29}^{+0.32}$ & two disc parameters hit limits, rel. nearby saturated star, very small part of the galaxy masked \\
     5  &  279324  &  E  & $0.070$ &  0.378 $\pm$ 0.007  & 3038$\pm$103 & 6.75$\pm$0.12  & AGN + S            & 8        &  $4.53_{-0.23}^{+0.29}$ &  8.86$_{-0.13}^{+0.15}$ & AGN hit constraint \\
     6  &  273518  &  SB & $0.075$ &  0.586 $\pm$ 0.015  & 3386$\pm$249 & 6.96$\pm$0.12  & AGN + S + D + B    & 17       &  $0.13_{-0.12}^{+0.34}$ &  4.00$_{-1.23}^{+3.33}$ & FAILED: $n$ hit lower limit \\
     7  &  238358  &  S  & $0.054$ &  1.630 $\pm$ 0.029  & 3313$\pm$35  & 7.11$\pm$0.10  & AGN + S + D        & 12       &  $2.63_{-0.13}^{+0.70}$ &  7.74$_{-0.15}^{+0.36}$ & spiral residuals, AGN hit constraint \\
     8  &  362697  &  E  & $0.054$ &  -  & - & -  & AGN + S            & 8        &  $3.23_{-0.03}^{+0.05}$ &  8.28$_{-0.08}^{+0.08}$ &  AGN hit constraint, poor PSF, strong central residuals \\
     9  &  240500  &  S0 & $0.075$ &  1.450 $\pm$ 0.017  & 3934$\pm$40  & 7.25$\pm$0.10  & AGN + S + D + star & 12 + 3   &  $1.63_{-0.16}^{+0.16}$ &  7.10$_{-0.21}^{+0.20}$  &  spheroid and disc similar in shape, elongated \\
    10  &  3631902 &  SB & $0.055$ &  3.090 $\pm$ 0.108  & 2916$\pm$135 & 7.20$\pm$0.10  & AGN + S + BE-D + B & 20       &  $0.19_{-0.18}^{+0.37}$ &  4.46$_{-1.24}^{+2.50}$  &  spheroid poorly constrained, two disc parameters hit limits\\
    11  &  609069  &  E  & $0.029$ &  -  & - & -  & AGN + S            & 8        &  $4.95_{-0.05}^{+0.05}$ &  9.01$_{-0.11}^{+0.11}$  &  central residuals, AGN hit constraint\\
    12  &  93676   &  E  & $0.095$ &  0.684 $\pm$ 0.007  & 1864$\pm$35  & 6.43$\pm$0.10  & AGN + S + galaxy & 8 + 7    &  $1.49_{-0.63}^{+0.63}$ &  6.95$_{-0.74}^{+0.74}$  & spheroid elongated \\
    13  &  619960  &  S0 & $0.080$ &  3.730 $\pm$ 0.030  & 1864$\pm$35  & 6.77$\pm$0.09  & AGN + S + D         & 12       &  $0.25_{-0.18}^{+0.22}$ &  3.88$_{-1.32}^{+1.54}$  & FAILED: galaxy masked, nearby saturated star, spheroid elongated \\
    14  &  163988  &  E  & $0.094$ &  0.786 $\pm$ 0.008  & 1795$\pm$35  & 6.43$\pm$0.10  & AGN + S            & 8        &  $1.18_{-0.07}^{+0.06}$ &  6.54$_{-0.19}^{+0.18}$ &  poor PSF, AGN hit constraint\\
    15  &  131124  &  E  & $0.091$ &  18.800$\pm$ 0.415  & 2554$\pm$34  & 7.39$\pm$0.09  & AGN + S            & 8        &  $1.00_{-0.19}^{+0.20}$ &  6.27$_{-0.37}^{+0.39}$ &  barely resolved\\
    16  &  178468  &  S0 & $0.091$ &  4.340 $\pm$ 0.117  & 2140$\pm$35  & 6.92$\pm$0.09  & AGN + S + D + galaxy & 12 + 7   &  $0.95_{-0.09}^{+0.09}$ &  6.18$_{-0.25}^{+0.25}$ &  system interacting/merging, rel. nearby saturated star, spheroid poorly constrained and elongated\\
    17  &  144673  &  E  & $0.083$ &  6.630 $\pm$ 0.054  & 2692$\pm$35  & 7.21$\pm$0.09  & AGN + S            & 8        &  $3.17_{-0.27}^{+0.34}$ &  8.25$_{-0.16}^{+0.20}$ & PA varies as function of radius?\\
    18  &  218688  &  E  & $0.087$ &  4.220 $\pm$ 0.021  & 2899$\pm$34  & 7.18$\pm$0.09  & AGN + S            & 8        &  $1.61_{-0.26}^{+0.26}$ &  7.08$_{-0.30}^{+0.31}$ & barely resolved \\
    19  &  348366  &  SB & $0.084$ &  1.690 $\pm$ 0.013  & 3520$\pm$69  & 7.18$\pm$0.10  & AGN + S + D + B + star & 17 + 3   &  $0.10_{-0.00}^{+0.00}$ &  3.70$_{-0.48}^{+0.48}$ & FAILED: $n$ hit lower limit \\
    20  &  311960  &  S  & $0.081$ &  1.460 $\pm$ 0.020  & 3796$\pm$69  & 7.22$\pm$0.10  & AGN + S + D + star+galaxy&12 + 10 &  $0.75_{-0.10}^{+0.10}$ &  6.18$_{-0.28}^{+0.28}$ &  rel. nearby saturated star, galaxy interacting and partly masked\\
    21  &  242362  &  S  & $0.077$ &  19.200$\pm$ 0.098  & 2692$\pm$35  & 7.41$\pm$0.09  & AGN + S + D        & 12       &  $1.35_{-0.10}^{+0.10}$ &  6.91$_{-0.20}^{+0.20}$ & spheroid elongated \\
    22  &  265000  &  S  & $0.068$ &  1.370 $\pm$ 0.031  & 1795$\pm$35  & 6.53$\pm$0.09  & AGN + S + D        & 12       &  $0.58_{-0.04}^{+0.04}$ &  5.87$_{-0.28}^{+0.28}$ &  poor PSF \\
    23  &  491383  &  SB & $0.076$ &  0.680 $\pm$ 0.013  & 2554$\pm$40  & 6.71$\pm$0.10  & AGN + S + D + B    & 17       &  $0.11_{-0.05}^{+0.08}$ &  3.81$_{-0.72}^{+1.04}$ & FAILED: $n$ hit lower limit \\
    24  &  508523  &  S0 & $0.090$ &  1.600 $\pm$ 0.016  & 2209$\pm$40  & 6.75$\pm$0.09  & AGN + S + D + star & 12 + 3   &  $0.68_{-0.67}^{+2.01}$ &  5.60$_{-1.71}^{+5.10}$ & spheroid very poorly constrained\\
    25  &  479971  &  SB & $0.082$ &  6.370 $\pm$ 0.672  & 2140$\pm$207 & 6.97$\pm$0.12  & AGN + S + D + B + star & 17 + 3   &  $1.36_{-0.37}^{+0.39}$ &  6.92$_{-0.38}^{+0.40}$ & spheroid faint, poor PSF\\
    26  &  492762  &  E  & $0.093$ &  16.500$\pm$ 0.433  & 2623$\pm$40  & 7.38$\pm$0.09  & AGN + S            & 8        &  $2.93_{-0.08}^{+0.09}$ &  8.11$_{-0.09}^{+0.10}$ & AGN hit constraint\\
    27  &  506001  &  S0 & $0.100$ &  1.960 $\pm$ 0.047  & 2002$\pm$35  & 6.70$\pm$0.09  & AGN + S + D        & 12       &  $0.29_{-0.14}^{+0.15}$ &  4.16$_{-0.90}^{+0.93}$ & spheroid elongated (fits lens?), visual classification uncertain \\
    28  &  521372  &  E  & $0.082$ &  14.500$\pm$ 0.040  & 3062$\pm$22  & 7.47$\pm$0.09  & AGN + S + star & 8 + 3    &  $2.15_{-0.04}^{+0.04}$ &  7.58$_{-0.10}^{+0.10}$ & central residuals, poor PSF, AGN hit constraint \\
     \hline
	\end{tabularx}
\end{table}
\end{landscape}

\subsection{SMBH mass comparison}
\label{sec:masses}

\begin{figure*}
\centering
    \includegraphics[width=\textwidth]{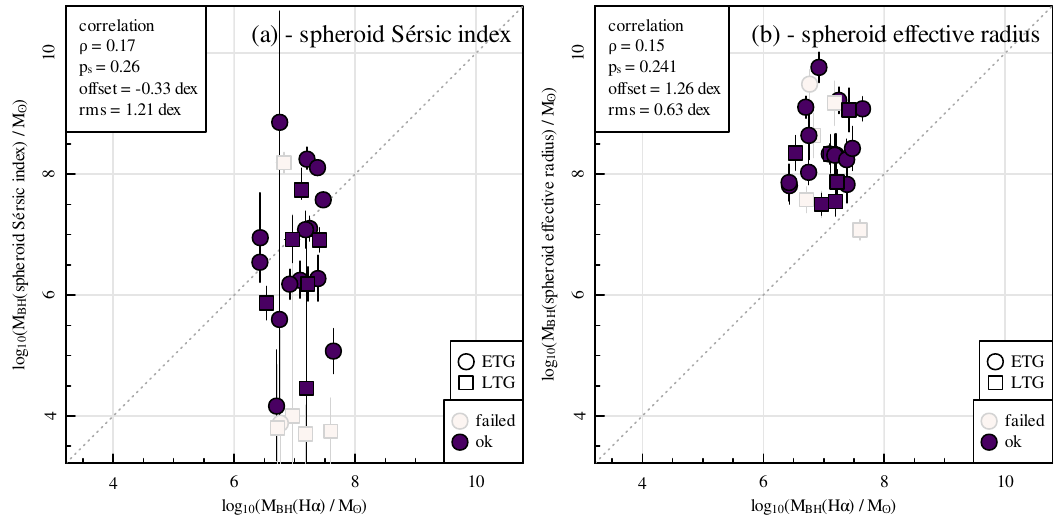}
    \includegraphics[width=\textwidth]{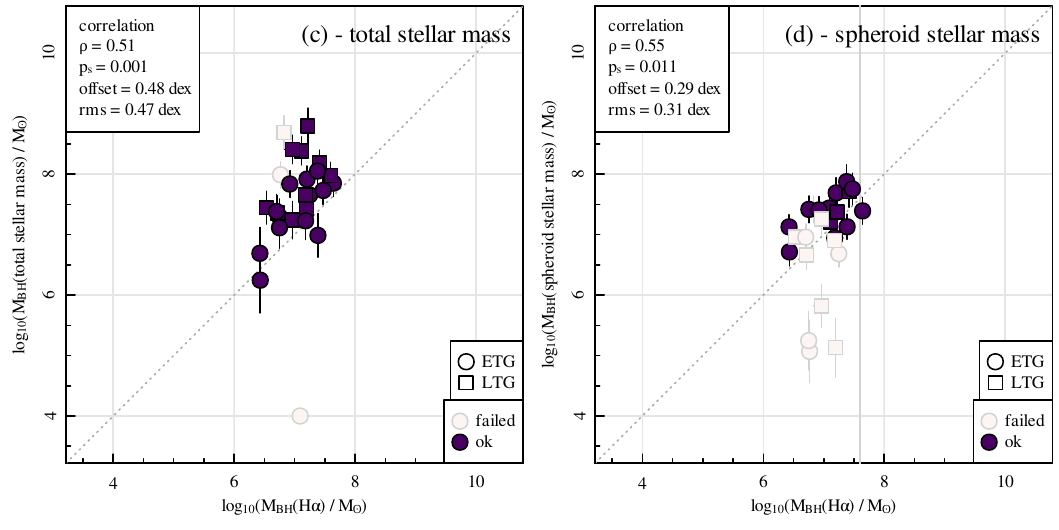}
	\caption{Comparison of the SMBH masses M$_{\mathrm{BH}}$(H$\alpha$)} obtained from the broad H$\alpha$ line  ($x$-axis in all cases) and four different photometry-based estimates ($y$-axes). \textbf{Top row:} M$_{\mathrm{BH}}$(spheroid S\'ersic index) (panel (a)) and M$_{\mathrm{BH}}$(spheroid effective radius) (panel (b)), obtained from the \citet{Sahu_2020} scaling relation between the SMBH mass and the spheroid S\'ersic index or the spheroid effective radius, respectively. \textbf{Bottom row:} M$_{\mathrm{BH}}$(total stellar mass) (panel (c)) and, M$_{\mathrm{BH}}$(spheroid stellar mass) (panel (b)), obtained from the \citet{Bentz2018} scaling relation between the SMBH mass and the galaxy stellar mass or bulge stellar mass, respectively, with the stellar mass estimated from the $g - i$ colour of our single S\'ersic fits or the spheroid component, respectively, in both cases using the calibration from \citet{Taylor2011}. Circles denote ETGs, squares denote LTGs, and different colors indicate the surface brightness fitting quality (cf. Section~\ref{sec:sbfittingresults}). Uncertainties on M$_{\mathrm{BH}}$(H$\alpha$) are shown, but typically smaller than the data points. The dashed line illustrates the one-to-one relation. The correlation coefficient $\rho$, the associated p-value p$_S$, the median offset from the 1:1-line and the offset-corrected rms scatter are indicated in the top left corner of each panel (in all cases excluding ``failed'' data points).
    \label{fig:mass}
\end{figure*}

Figure~\ref{fig:mass} shows the main results of this work: the four different versions of the photometric SMBH mass estimates compared against the spectroscopic measurements in each case. Table~\ref{tab:masses} provides the corresponding numerical values. The top row focuses on the SMBH masses derived from the \citet{Sahu_2020} relations (equations~\ref{eq:SahurelationETG} to~\ref{eq:SahuReLTG}), based on the spheroid S\'ersic index (panel (a)) and spheroid effective radius (panel (b)). The bottom row shows the SMBH masses derived from the stellar mass of the galaxy (panel (c)) or bulge (panel (d)) with the scaling relations presented in \citet[equations~\ref{eq:Bentz2018Mbulge} and~\ref{eq:Bentz2018Mstar}]{Bentz2018}. In each panel, we show the photometric mass estimate with uncertainties against the H$\alpha$-based masses with uncertainties (typically smaller than the data points), colour coded by the surface brightness fitting success, and indicating the correlation coefficient $\rho$, the associated p-value p$_S$, the median offset from the 1:1-line and the offset-corrected rms scatter, always excluding ``failed'' data points.

We quote the weighted correlation coefficient ($\rho$), calculated using the \textsc{stats} package of the \textsc{R} programming language \citep{R2024}, which tests for a linear relation taking the measurement uncertainties of both mass estimates into account. We obtain consistent results when using the Pearson correlation coefficient instead. The associated p-value of significance, p$_S$, is obtained from the tail of values falling below zero when bootstrapping the sample 10\,000 times. Again, we obtain very similar results when using a permutation analysis instead, where the p-value is defined as the fraction of values falling below the measured one.

It becomes clear from Figure~\ref{fig:mass} that both mass estimates derived from the \citet{Sahu_2020} relations (top row) do not correlate with the H$\alpha$-based masses: the scatter is large, correlation coefficients are low ($<\,0.2$) and p$_S>\,0.05$. Panel (b), i.e. M$_{\mathrm{BH}}$(spheroid effective radius), also shows a significant positive offset. The SMBH mass estimates based on stellar mass (bottom row) yield significant results with p$_S<\,0.05$, correlation coefficents $>$\,0.5 and significantly smaller scatter. Uncertainties on the photometric mass estimates are generally larger than those on the spectroscopic ones.\\

The highest correlation with the H$\alpha$ based SMBH masses is achieved for M$_{\mathrm{BH}}$(spheroid stellar mass) (panel (d) in Figure~\ref{fig:mass}). This is obtained from the bulge stellar mass - SMBH mass relation for active galaxies from \citet[equation~\ref{eq:Bentz2018Mbulge}]{Bentz2018}, with M$_{*, \mathrm{bulge}}$ derived from the $g-i$ colour of the spheroid component with the \citet{Taylor2011} calibrations. In panel (c), we see a similarly strong correlation - albeit with more scatter - between M$_{\mathrm{BH}}$(H$\alpha$) and M$_{\mathrm{BH}}$(total stellar mass), which, instead of the spheroid mass, uses the total galaxy stellar mass based on the $g - i$ colour of our automated single S\'ersic fits. This correlation is expected to show a higher scatter
especially for galaxies containing discs (LTG and S0), which contaminate the the total galaxy stellar mass \citep[e.g.][]{2013ARA&A..51..511K}. We also observe a positive offset which is comparable to the intrinsic scatter. The most likely cause for this is AGN contamination, since the \citet{Taylor2011} relations we use for converting colour to stellar mass are not tailored to type 1 AGN (see Section~\ref{sec:scalingrelations}); and thus the presence of additional flux from the AGN would generally lead to a mass overestimate. Separating the AGN and spheroid components succeeds in reducing both the offset and the scatter, as evidenced by panel (d). \\

In contrast to M$_{\mathrm{BH}}$(spheroid stellar mass) (panel (d) of Figure~\ref{fig:mass}), neither M$_{\mathrm{BH}}$(spheroid S\'ersic index) (panel (a) in Figure~\ref{fig:mass}) nor M$_{\mathrm{BH}}$(spheroid effective radius) (panel (b) in Figure~\ref{fig:mass}) show a significant correlation with M$_{\mathrm{BH}}$(H$\alpha$), despite being based on the same fits (just using the spheroid S\'ersic index and spheroid effective radius instead of the spheroid magnitude). The uncertainties on $n$ and $R_\mathrm{e}$ are typically higher than those on $m$ (cf. Section~\ref{sec:errorestimation}), which is reflected in the larger error bars in panels (a) and (b) of Figure~\ref{fig:mass}, but even taking these into account we do not obtain a significant correlation.

We remind the reader that the \citet{Sahu_2020} relations upon which the top row of Figure~\ref{fig:mass} is based, are derived from 1-dimensional surface brightness fits to space-based infrared data (3.6\,$\mu$m; see Section~\ref{sec:scalingrelations}); and not specifically tailored to be applied to objects hosting an AGN. From the present study it seems that they are not directly applicable to the type of data (ground-based seeing-limited optical data close to the resolution limit), object (type 1 AGN) or analysis method (two-dimensional decompositions) used here. Alternatively, it may be that there is an underlying correlation in panels (a) and (b) of Figure~\ref{fig:mass}, but the scatter is too large to observe it for a number of reasons: first, these relations use only a single photometric band ($i$-band); second, $n$ and $R_\mathrm{e}$ are typically less robustly determined than $m$; third, our sample size is small; and fourth, the scaling relations have an intrinsic scatter which is comparable to or larger than most of the uncertainty estimates of our values -- \citet{Sahu_2020} quote a total rms scatter of 0.65 and 0.67 for ETGs and LTGs, respectively. The strong offset in the case of M$_{\mathrm{BH}}$(spheroid effective radius) is most easily explained by the different wavelength ranges used.

We provide a more detailed account of some of the most notable outliers and trends observed in panel (a) of Figure~\ref{fig:mass} in Appendix~\ref{app:discusssahu}.

\section{Conclusions}
\label{sec:conclusion}

In this work we performed a case study on estimating the SMBH mass of 28 type 1 AGN located in nearby galaxies, selected from the GAMA survey with both spectroscopic and photometric data available. We aim to test the applicability of the \citet{Sahu_2020} M$_{\mathrm{BH}}$ - $n$ relation to ground-based optical data and type 1 AGN; and compare results to more traditional methods of estimating SMBH masses from photometric data.

We performed careful spectral decomposition to extract the pure AGN emission from the spectra greatly dominated by the host emission and calculate the mass of the SMBH using its widely used scaling relation with the broad H$\alpha$ line luminosity and width. We use these spectral SMBH mass estimates (M$_{\mathrm{BH}}$(H$\alpha$)) as reference values, with which we compare four different variants of photometric SMBH mass estimates derived from an independent analysis of the imaging data. For this, we perform detailed surface brightness decompositions of the galaxy images in the optical $g$ and $i$ bands in order to decouple the spheroid contribution from the AGN emission and remaining host galaxy. We then obtain:
\begin{enumerate}
    \item M$_{\mathrm{BH}}$(spheroid S\'ersic index), 
    derived from the spheroid S\'ersic index with the M$_\mathrm{BH}$~-~$n$ relation from \citet[panel (a) in Figure~\ref{fig:mass}]{Sahu_2020};
    \item M$_{\mathrm{BH}}$(spheroid effective radius), 
    derived from the spheroid effective radius with the M$_\mathrm{BH}$~-~$R_\mathrm{e}$ relation from \citet[panel (b) in Figure~\ref{fig:mass}]{Sahu_2020};
    \item M$_{\mathrm{BH}}$(total stellar mass),
    derived from the total stellar mass of the galaxy with the M$_\mathrm{BH}$~-~M$_{*, \mathrm{tot}}$ relation from \citet{Bentz2018}, 
    where the stellar mass is obtained from the $g-i$ colour of our single S\'ersic fits using the calibration by \citet[panel (c) in Figure~\ref{fig:mass}]{Taylor2011}
    \item M$_{\mathrm{BH}}$(spheroid stellar mass),
    derived from the spheroid stellar mass with the M$_\mathrm{BH}$~-~M$_{*, \mathrm{bulge}}$ relation from \citet{Bentz2018}, 
    where the spheroid mass is obtained from the $g - i$ colour of our multi-component decompositions using the calibration by \citet[panel (d) in Figure~\ref{fig:mass}]{Taylor2011}.
\end{enumerate}

We find that the SMBH mass estimates based on the $i$-band spheroid S\'ersic index or effective radius, do not correlate with the spectroscopic SMBH masses. This shows that the M$_\mathrm{BH}$~-~$n$ and M$_\mathrm{BH}$~-~$R_\mathrm{e}$ relations are not applicable to estimate the SMBH mass in a pure AGN sample from ground-based optical imaging. 

SMBH mass estimates within $\sim$~0.3\,dex of the corresponding mass estimates measured from spectroscopic data can be obtained with as few as two optical wavelength bands through a careful multi-component modelling of the surface brightness distribution and a subsequent estimate of the spheroid stellar mass based on its $g - i$ colour. However, the success of such an analysis is sensitive to the quality of the data and the host galaxy type (i.e. its number of stellar components), and is computationally intensive. For automated analyses of large samples of galaxies in surveys, an additional challenge will be to define the number of components to fit for each galaxy; and to assess the robustness of the results (for example against component swapping). A less computationally expensive route which can easily be automated for large samples of galaxies is to use the total galaxy mass estimated from $g$ and $i$ band single S\'ersic fits, but this comes at the cost of higher scatter ($\sim$~0.5\,dex) and a higher probability of systematic biases (e.g. due to AGN contamination) in the derived SMBH masses.

In addition to the methods tested in this exploratory study, SED fitting is widely used to obtain stellar masses of galaxies. Whilst this is not the focus of the present work, we have confirmed that using the SED-based masses from the GAMA database (\textsc{StellarMassesGKVv24}, which is the most recent update of the \citealt{Taylor2011} stellar mass catalogue) yields SMBH masses which correlate with the H$\alpha$-based masses on a level comparable to that based on our own total stellar mass estimates. To reduce the scatter in SMBH mass measurements and obtain reliable \emph{spheroid} stellar masses, the simultaneous SED and surface brightness fitting of multiple wavelength bands, e.g. using \textsc{ProFuse} \citep{Robotham2022}, may provide a promising route. To aid the analysis of extremely large datasets, machine learning tools might be used.\\

We conclude that:
\begin{enumerate}
    \item the complex spectral fittings, even if optical spectra are of poor S/N ratio or with significant stellar contribution from the host galaxy, were successful in extracting the AGN contribution; the broad H$\alpha$ line was detected and its parameters (luminosity and width) measured in 26 AGN (Section~\ref{sec:spectralfittings});
    \item deriving reliable spheroid S\'ersic indices and effective radii from the surface brightness decompositions proved to be challenging and large uncertainties remain, mostly due to degeneracies between model parameters of the AGN and the different stellar components of the host galaxy, which cannot be sufficiently constrained given the data quality at hand, especially its resolution (Section~\ref{sec:sbfittingresults}); 
    \item combined with the intrinsically large scatter in the \citet{Sahu_2020} relations and its uncertain applicability range, this yields the method unsuitable for SMBH mass estimation in our sample (Section~\ref{sec:masses});
    \item instead, a more robust SMBH mass can be obtained from a rough estimate of stellar mass in the spheroid or even in the galaxy as a whole (Section~\ref{sec:masses}).
 \end{enumerate}   

In view of present and forthcoming high quality large imaging surveys like Euclid or LSST, we thus believe that the most viable method to estimate SMBH masses for large numbers of AGN will use the well-established scaling relations between SMBH mass and galaxy or spheroid stellar mass. Depending on the data quality (depth and resolution), number of photometric bands available, computational resources, required level of automisation and desired level of accuracy of the SMBH masses, the optimal strategy to estimate stellar mass may vary. It is advisable to perform a small pilot study first in which several methods are tested and compared, ideally against independent data e.g. from overlapping spectroscopy. The findings of this work may serve as a guide in performing such studies, in order to obtain reliable SMBH masses for large samples of AGN in modern imaging surveys.

\section*{Acknowledgements}

The first two authors should be regarded as joint first authors. Co-first authors can prioritise their names when adding this paper’s reference to their r\'esum\'es.

We thank the anonymous referee for their very detailed and constructive feedback which truly and substantially improved this manuscript.

GAMA is a joint European-Australasian project based around a spectroscopic campaign using the Anglo-Australian Telescope. The GAMA input catalogue is based on data taken from the Sloan Digital Sky Survey and the UKIRT Infrared Deep Sky Survey. Complementary imaging of the GAMA regions is being obtained by a number of independent survey programmes including GALEX MIS, VST KiDS, VISTA VIKING, WISE, Herschel-ATLAS, GMRT, and ASKAP providing UV to radio coverage. GAMA was funded by the Science and Technology Facilities Council (STFC), the Australian Research Council (ARC), the Anglo-Australian Observatory (AAO), and the participating institutions. The GAMA website is https://www.gama-survey.org/.

D. Ili\'c. acknowledges funding provided by the University of Belgrade - Faculty of Mathematics (the contract \textnumero451-03-47/2023-01/200104) through the grant of the Ministry of Science, Technological Development and Innovation of the Republic of Serbia, and the support of the Alexander von Humboldt Foundation. S. Casura acknowledges support by the Claussen-Simon-Stiftung in Hamburg. J. Liske acknowledges support by the Deutsche Forschungsgemeinschaft (DFG) under Germany’s Excellence Strategy – EXC 2121 ``Quantum Universe'' – 390833306.

Based on observations made with ESO Telescopes at the La Silla Paranal Observatory under programme IDs 177.A-3016, 177.A- 3017, 177.A-3018, and 179.A-2004, and on data products produced by the KiDS consortium. The KiDS production team acknowledges support from: Deutsche Forschungsgemeinschaft (DFG), European Research Council (ERC), Netherlands Research School for Astronomy (NOVA), and Dutch Research Council (NWO) M grants; Target; the University of Padova, and the University Federico II (Naples).

\section*{Data Availability}

The authors confirm that the data supporting the findings of this study are available within the article. The data used in this study are accessible from the GAMA\footnote{https://www.gama-survey.org} online database.



\bibliographystyle{mnras}
\bibliography{dilic_references,zzzreferences} 



\appendix

\section{Derivation of initial guesses}
\label{app:initialguesses}

We obtain initial guesses for all parameters used in the surface brightness fitting (Section~\ref{sec:sbfitting}) from the \textsc{BDDecomp} DMU pipeline (see Section~\ref{sec:datasample}), using the single S\'ersic fit results for elliptical galaxies and the S\'ersic plus exponential fits for lenticular and (barred) spiral galaxies. Since AGN and bar components were not considered in \textsc{BDDecomp}, we simply split the spheroid flux evenly between all of those central components, while leaving disc magnitudes unchanged. Broken exponential discs have the starting values of $h_1$ and $R_\mathrm{b}$ set to $R_\mathrm{e}$ of the (non-broken) exponential disc fitted in \textsc{BDDecomp}; while $h_2$ takes half of this value and $a$ is set to 0.2. Ferrers bars take the $R_\mathrm{out}$ initial guess from $R_\mathrm{e}$, with $\alpha$\,=\,1.5 and the position angle set to zero. The spheroid S\'ersic index was always set to 4 initially. All other parameters are taken directly from the automated fits. 

Since nearby sources were also not considered in the \textsc{BDDecomp} pipeline, we use the manual definitions of their approximate position as initial guesses for $x$ and $y$, while the magnitude is set equal to the spheroid magnitude and - for those objects fitted with S\'ersic functions - the initial values of $R_\mathrm{e}$, $n$, $b/a$ and PA are set to 5\,pix, 1, 1 and $0\deg$ respectively. Since these are rather arbitrary, we run a short optimisation of the nearby source parameters only, using only its own segment for fitting before adding the components (and segment) of the main galaxy for joint fitting. This leads to a faster convergence of the much more complex joint fit, thus decreasing the overall run-time despite the extra fitting step. 

After the first fitting round, we improved the initial guesses as necessary to avoid the swapping of components (cf. Section~\ref{sec:initialguesses}): 
\begin{itemize}
\item If the point source was faint or discarded entirely (before we implemented the constraint forcing it to be significant; cf. Section~\ref{sec:fitconstraints}), we split the spheroid magnitude giving two thirds to the point source and one third to the spheroid; and set the corresponding S\'ersic index to a maximum of 4.
\item If the bulge and disc components were swapped, we exchanged all parameters except S\'ersic index, which was set to 4 (this is the procedure applied in the automated \textsc{BDDecomp} pipeline, see \citealt{Casura2022}).
\item If the bar was swapped with one of the other components, or all three components were swapped, we exchanged the values fitted for $b/a$, PA and the magnitude; estimated $R_\mathrm{out}$ from the diagnostic plot, set $a\,=\,1.5$ and inserted missing bulge and disc parameters as needed, using values of $n\,=\,4$, $R_\mathrm{e}\,=\,R_\mathrm{out}$ for the disc or $R_\mathrm{e}\,=\,2$\,pix for the bulge, $b/a$\,=\,1 and PA\,=\,0 or 90\,deg depending on the orientation of the bar (preferably choosing a value perpendicular to the PA of the bar in order to avoid bulge-bar swapping). 
\item In one case, the bar component was discarded; here we initiated it with one third of the disc magnitude (reducing the latter accordingly), the disc position angle, $b/a\,=\,0.2$, $a\,=\,1.5$ and $R_\mathrm{out}$ estimated from the fit diagnostic plot.
\item Additionally, if the bulge axial ratio was below 0.6, we set it back to 1 (round) in order to limit bulge-bar degeneracies. 
\end{itemize}

This procedure was successful in ``unswapping'' all components for our 28 galaxies; and also recovered the bar for the one case where it had been discarded. However, it did not recover any of the discarded point sources; nor result in rounder bulges than earlier fits. This is because in these cases, the likelihood space is not characterised by two (or more) strong, far-apart peaks as is the case for swapped components. Thus, since our routine is only weakly dependent on initial guesses, changing the initial guess did not influence the fit in these cases. For this reason, we introduced the explicit constraint on the brightness of the AGN as described in Section~\ref{sec:fitconstraints}. Elongated bulges were accepted as reasonable fits as long as they were not aligned with the bar component (where present), but flagged for further inspection.

\section{Corner plots}
\label{app:cornerplots}
\begin{figure*}
\centering
    \includegraphics[width=\textwidth]{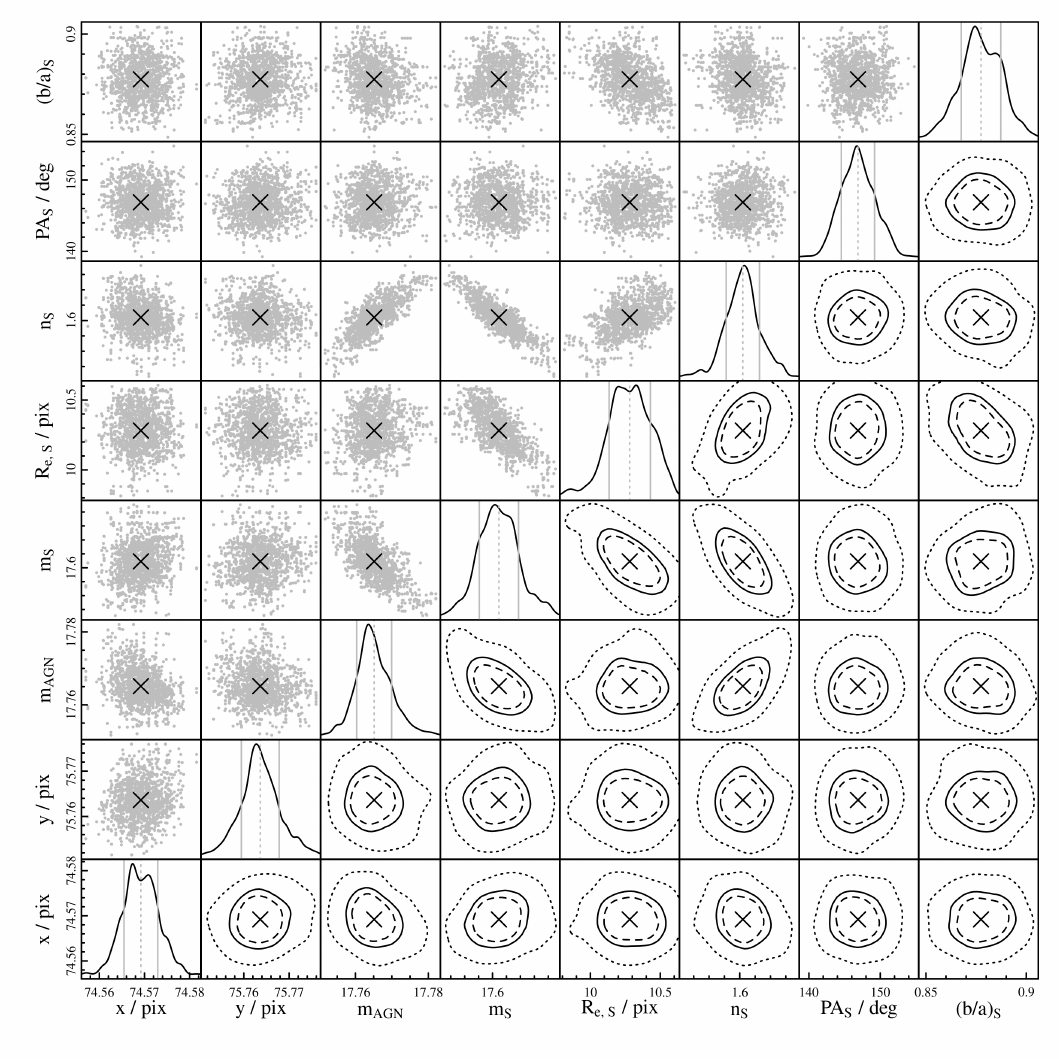}
    \caption{The corner plot of the surface brightness fit to galaxy 218688 (shown in Figure~\ref{fig:examplefitE}), classified as elliptical (8 free fitting parameters).
    For each pair of parameters, we show the distribution of the MCMC chain points in the upper left half of the plot; with corresponding contours including 50\,\%, 68\,\% and 95\,\% of the data (dashed, solid and dotted lines respectively) in the lower right set of panels. Crosses indicate the mean of each distribution and panels on the diagonal show the one-dimensional distribution for each parameter with the mean and 1$\sigma$ uncertainty range indicated. The meaning of each parameter is described in Section~\ref{sec:fitconstraints}. Those labelled ``AGN'' refer to the AGN component, ``S'' is the spheroid. The AGN magnitude, spheroid magnitude, S\'ersic index and effective radius are correlated (cf. Section~\ref{sec:errorestimation}); while the position, and spheroid position angle and axial ratio are independent.
    }
    \label{fig:cornerplotE} 
\end{figure*}

\begin{figure*}
\centering
    \includegraphics[width=\textwidth]{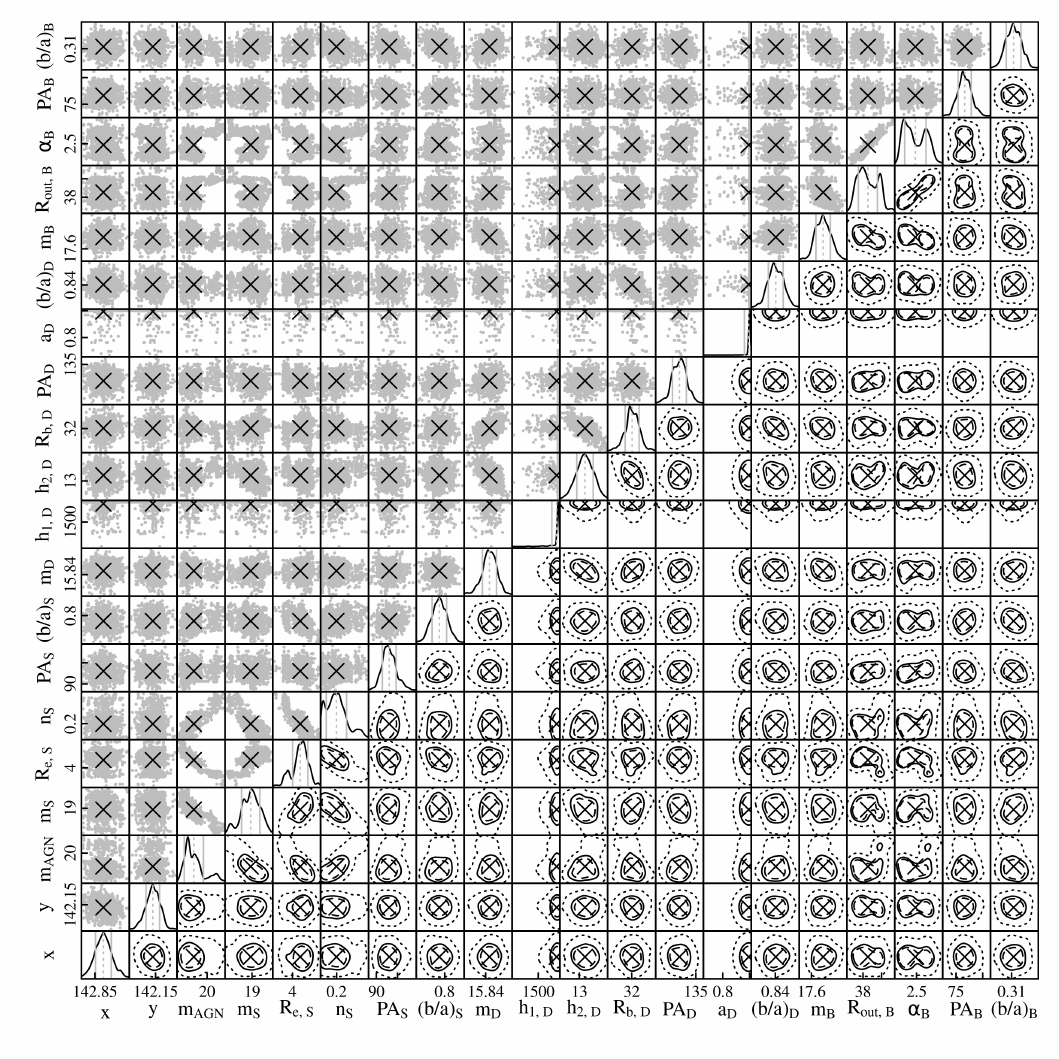}
    \caption{The corner plot of the surface brightness fit to galaxy 3631902 (shown in Figure~\ref{fig:examplefitSB}), classified as a barred spiral galaxy with a broken exponential disc (20 free fitting parameters). The plot layout is the same as in Figure~\ref{fig:cornerplotE}. The meaning of each parameter is described in Section~\ref{sec:fitconstraints}. Those labelled ``AGN'' refer to the AGN component, ``S'' is the spheroid, ``D'' the (broken exponential) disc, and ``B'' the bar. For readability, units are omitted from this figure: positions, scale lengths and radii are given in pix; angles in degrees. The broken exponential parameters $h_1$ (inner scale length) and $a$ (sharpness of transition) both hit their upper limits; and several other parameters show degeneracies (see discussion in Section~\ref{sec:errorestimation}).}
    \label{fig:cornerplotSB} 
\end{figure*}

Figures~\ref{fig:cornerplotE} and~\ref{fig:cornerplotSB} show corner plots for the surface brightness fits to the two example galaxies presented in Figures~\ref{fig:examplefitE} and ~\ref{fig:examplefitSB}. Essentially, these are slices through the N-dimensional likelihood space, showing the distribution of likelihood samples for each pair of parameters. We consider only the final batch of 2000 samples returned by CHARM (cf. Section~\ref{sec:convergence}) in producing these plots.

\section{SMBH mass estimates}
\label{app:masstable}
Table~\ref{tab:masses} lists all SMBH mass estimates derived in this work. 
\begin{table*}
	\centering
	\caption{All SMBH mass estimates derived in this work for our sample of 28 galaxies, see Section~\ref{sec:errorestimation} for details. For readability, and since the uncertainties tend to be symmetric, we do not quote separate lower and upper limits. Values with a * indicate failed fits (see Table~\ref{tab:sbfittingresults} and Section~\ref{sec:sbfittingresults}). Two values of M$_{\mathrm{BH}}$(H$\alpha$) are missing since they do not contain type 1 AGN (Section~\ref{sec:spectralfittings}). One value of M$_{\mathrm{BH}}$(total stellar mass) is missing since the galaxy failed in the automated single S\'ersic fits.}
	\label{tab:masses}
	\begin{tabular}{rrccccc}
    & & log$_{10}$(M$_{\mathrm{BH}}$ & log$_{10}$(M$_{\mathrm{BH}}$ & log$_{10}$(M$_{\mathrm{BH}}$ & log$_{10}$(M$_{\mathrm{BH}}$ & log$_{10}$(M$_{\mathrm{BH}}$ \\
 	ID & \textsc{CATAID} & (H$\alpha$)) & (spheroid S\'ersic index)) & (spheroid effective radius)) & (total stellar mass)) & (spheroid stellar mass)) \\ 
    & & [M$_{\odot}$] & [M$_{\odot}$] & [M$_{\odot}$] & [M$_{\odot}$] & [M$_{\odot}$]\\
    \hline
      1  & 23104  & 7.64 $\pm$ 0.09 & 5.07 $\pm$ 0.37 & 9.08 $\pm$ 0.20 & 7.85 $\pm$ 0.22 & 7.39 $\pm$ 0.22 \\
      2  & 47558  & 7.60 $\pm$ 0.10 & *3.75 $\pm$ 0.55 & *7.08 $\pm$ 0.17 & 7.98 $\pm$ 0.22 & ~*$10^6$ $\pm$ $10^7~$ \\
      3  & 323124 & 6.83 $\pm$ 0.12 & *8.19 $\pm$ 0.17 & *8.64 $\pm$ 0.37 & *8.69 $\pm$ 0.28 & *7.42 $\pm$ 0.22 \\
      4  & 278841 & 7.09 $\pm$ 0.09 & 6.24 $\pm$ 0.31 & 8.33 $\pm$ 0.17 & - & 7.44 $\pm$ 0.22 \\
      5  & 279324 & 6.75 $\pm$ 0.12 & 8.86 $\pm$ 0.14 & 8.64 $\pm$ 0.40 & 7.29 $\pm$ 0.30 & 7.42 $\pm$ 0.22 \\
      6  & 273518 & 6.96 $\pm$ 0.12 & *4.00 $\pm$ 2.28 & *7.50 $\pm$ 0.21 & 7.24 $\pm$ 0.31 & *5.82 $\pm$ 0.36 \\
      7  & 238358 & 7.11 $\pm$ 0.10 & 7.74 $\pm$ 0.26 & 8.33 $\pm$ 0.32 & 8.39 $\pm$ 0.23 & 7.21 $\pm$ 0.21 \\
      8  & 362697 & - & 8.28 $\pm$ 0.08 & 8.87 $\pm$ 0.45 & 8.52 $\pm$ 0.25 & 8.27 $\pm$ 0.35 \\
      9  & 240500 & 7.25 $\pm$ 0.10 & 7.10 $\pm$ 0.21 & 9.22 $\pm$ 0.22 & 7.66 $\pm$ 0.24 & *6.68 $\pm$ 0.22 \\
      10 & 3631902& 7.20 $\pm$ 0.10 & 4.46 $\pm$ 1.87 & 7.55 $\pm$ 0.24 & 7.42 $\pm$ 0.28 & *5.13 $\pm$ 0.48 \\
      11 & 609069 & - & 9.01 $\pm$ 0.11 & 8.28 $\pm$ 0.43 & 7.16 $\pm$ 0.32 & 7.38 $\pm$ 0.22 \\
      12 & 93676  & 6.43 $\pm$ 0.10 & 6.95 $\pm$ 0.74 & 7.81 $\pm$ 0.31 & 6.25 $\pm$ 0.54 & 6.71 $\pm$ 0.23 \\
      13 & 619960 & 6.77 $\pm$ 0.09 & *3.88 $\pm$ 1.43 & *9.49 $\pm$ 0.24 & *7.99 $\pm$ 0.21 & *5.07 $\pm$ 0.52 \\
      14 & 163988 & 6.43 $\pm$ 0.10 & 6.54 $\pm$ 0.19 & 7.86 $\pm$ 0.30 & 6.69 $\pm$ 0.43 & 7.13 $\pm$ 0.20 \\
      15 & 131124 & 7.39 $\pm$ 0.09 & 6.27 $\pm$ 0.38 & 7.83 $\pm$ 0.30 & 6.99 $\pm$ 0.36 & 7.13 $\pm$ 0.21 \\
      16 & 178468 & 6.92 $\pm$ 0.09 & 6.18 $\pm$ 0.25 & 9.76 $\pm$ 0.25 & 7.84 $\pm$ 0.22 & 7.41 $\pm$ 0.22 \\
      17 & 144673 & 7.21 $\pm$ 0.09 & 8.25 $\pm$ 0.18 & 8.32 $\pm$ 0.35 & 7.92 $\pm$ 0.22 & 7.69 $\pm$ 0.25 \\
      18 & 218688 & 7.18 $\pm$ 0.09 & 7.08 $\pm$ 0.30 & 8.31 $\pm$ 0.35 & 7.23 $\pm$ 0.31 & 6.94 $\pm$ 0.21 \\
      19 & 348366 & 7.18 $\pm$ 0.10 & *3.70 $\pm$ 0.48 & *9.17 $\pm$ 0.37 & 7.66 $\pm$ 0.24 & *6.90 $\pm$ 0.23 \\
      20 & 311960 & 7.22 $\pm$ 0.10 & 6.18 $\pm$ 0.28 & 7.87 $\pm$ 0.21 & 8.80 $\pm$ 0.29 & 7.37 $\pm$ 0.22 \\
      21 & 242362 & 7.41 $\pm$ 0.09 & 6.91 $\pm$ 0.20 & 9.06 $\pm$ 0.36 & 8.18 $\pm$ 0.22 & 7.71 $\pm$ 0.26 \\
      22 & 265000 & 6.53 $\pm$ 0.09 & 5.87 $\pm$ 0.28 & 8.35 $\pm$ 0.29 & 7.44 $\pm$ 0.27 & *6.96 $\pm$ 0.20 \\
      23 & 491383 & 6.71 $\pm$ 0.10 & *3.81 $\pm$ 0.88 & *7.58 $\pm$ 0.21 & 7.36 $\pm$ 0.29 & *6.66 $\pm$ 0.24 \\
      24 & 508523 & 6.75 $\pm$ 0.09 & 5.60 $\pm$ 3.40 & 8.03 $\pm$ 0.20 & 7.11 $\pm$ 0.34 & *5.25 $\pm$ 0.48 \\
      25 & 479971 & 6.97 $\pm$ 0.12 & 6.92 $\pm$ 0.39 & 7.50 $\pm$ 0.18 & 8.41 $\pm$ 0.24 & *7.26 $\pm$ 0.26 \\
      26 & 492762 & 7.38 $\pm$ 0.09 & 8.11 $\pm$ 0.09 & 8.24 $\pm$ 0.34 & 8.05 $\pm$ 0.22 & 7.88 $\pm$ 0.28 \\
      27 & 506001 & 6.70 $\pm$ 0.09 & 4.16 $\pm$ 0.92 & 9.11 $\pm$ 0.18 & 7.38 $\pm$ 0.28 & *6.96 $\pm$ 0.22 \\
      28 & 521372 & 7.47 $\pm$ 0.09 & 7.58 $\pm$ 0.10 & 8.42 $\pm$ 0.36 & 7.73 $\pm$ 0.23 & 7.75 $\pm$ 0.26 \\
     \hline
	\end{tabular}
\end{table*}

\section{Discussion of Figure~4, panel (a)}
\label{app:discusssahu}

Since testing the applicability of the M$_{\mathrm{BH}}$ - $n$ relation for our type of data, object and analysis method was the primary motivation for performing this work, we discuss the results shown in panel (a) of Figure~7 in more detail here. For more general comments on the lack of a correlation between M$_{\mathrm{BH}}$(spheroid S\'ersic index) and M$_{\mathrm{BH}}$(H$\alpha$), see Section~\ref{sec:masses}. In addition to the reasons listed there, a number of systematic uncertainties remaining in our surface brightness analysis (cf. Section~\ref{sec:sbfittingresults}) may influence the S\'ersic index based SMBH mass estimates.

For example, the outlier at the bottom left of the figure is galaxy 23104, which has the comment ``spheroid elongated'' (cf. Table~\ref{tab:sbfittingresults}). Given that its S\'ersic index based mass estimate is considerable too low, it seems likely that the S\'ersic component did not fit an actual spheroid but instead a lens, embedded disc, or similar non-spheroidal central component, which is difficult to identify due to the resolution limit of the data. Of the seven objects that have the comment ``spheroid elongated'' in Table~\ref{tab:sbfittingresults}, all but one fall below the 1:1-line, indicating that their S\'ersic indices are generally too small when taking the H$\alpha$-estimated SMBH mass as a reference. Potentially, thus, spheroid elongation can be used as an additional criterion for the assessment of fit qualities. 

A notable outlier with a very high M$_{\mathrm{BH, n}}$ estimate is galaxy 279324. This object has a noisy spectrum with the possibility that the H$\alpha$-derived BH mass is inaccurate, due to poor estimate of the underlying continuum.
The surface brightness fitting has the comment ``AGN hit constraint'', meaning that the brightness of the AGN component needed to be forced (see Section~\ref{sec:fitconstraints}). The choice of how bright to force the AGN component influences the derived spheroid S\'ersic index, and is somewhat arbitrary. Our choice was guided by visual inspection of the fit results, but the exact flux allocation to the different components is difficult based on visual inspection alone; and a slightly stronger constraint - resulting in correspondingly lower S\'ersic index based mass estimates for the galaxies affected by this constraint - would be justifiable, too. In comparison to the spectral fitting, indeed all eight galaxies which hit the AGN constraint lie above the 1:1-line. This indicates that we could have obtained more consistent results by implementing a stronger constraint.

Other categories of galaxies with comments in Table~\ref{tab:sbfittingresults} also show trends with position in panel (a) of Figure~\ref{fig:mass}. Objects that generally fall below the 1:1-line (like those with elongated spheroids) have underestimated spheroid S\'ersic indices, which can be an indication that discy or other non-spheroidal components are fitted. This is the case for all objects with nearby saturated stars and/or interacting systems, the two galaxies with broken exponential discs, the two barely resolved objects and one object where we suspect a position angle varying as a function of radius. The former of these categories have low fit qualities; while the latter suffer from low number statistics. 
Objects that fall above the 1:1 line have overestimated S\'ersic indices, which most often indicates that the S\'ersic function is accounting for some of the AGN flux, i.e. the AGN component was not fully subtracted by the point source function. Apart from the galaxies that hit their AGN constraint, this is also the case for all galaxies which show strong residuals at their centres. Objects flagged as having particularly poor PSF estimates, however, do not show any obvious pattern in their distribution in panel (a) of Figure~\ref{fig:mass}. \\

A better agreement for individual estimates could be achieved by tuning the surface brightness fitting parameters accordingly (e.g., the AGN brightness constraint), or by considering more parameters in assessing the fit reliability (e.g., spheroid elongation). However, we do not add such improvements, since they build upon the independent mass estimate being available from the spectral fitting. In our current analysis, the spectral fitting and the surface brightness fitting were treated entirely independently, using different data and lead by two different people (DI and SC respectively). As such, this blind analysis is one of the main strengths of this work.


\bsp	
\label{lastpage}
\end{document}